\documentclass[preprint,preprintnumber]{revtex4}

\usepackage{graphicx}

\usepackage{amssymb}

\usepackage{mathrsfs}

\begin{document}

\title{Topological Insulators ,Weyl Semimetals  and   Topological Superconductors  -A Transport View }

\author{D. Schmeltzer}

\affiliation{Physics Department, City College of the City University of New York,  
New York, New York 10031, USA}

\pacs{}

\begin{abstract}

We show that the  electronic band  theory in the momentum space requires  information about the  transport of the eigenfunctions . The  transport of the  eigenfunctions in the Brillouine Zone induces spin connections (gauge field in the momentum space) and curvatures (the equivalent of the electromagnetic strength). When the theory is applied to topological materials  characterized by discrete symmetries   such as time reversal,parity inversion, and charge conjugation the curvature and the spin connections  needs to satisfy  constraints  conditions. As a result of the constraints the curvature generates   topological invariants such as the first and second Chern number. The Chern numbers are reveald  by measuring  the response to an   external electromagnetic  field. 

\noindent 
We will  study transport in  Topological Insulators,Weyl Semimetals (on flat and curved surfaces) and Topological Superconductors  for which  we  compute the Andreev    crossed reflection. Using a $p-wave$ wire coupled to two rings  we  show that  the persistent current in the   rings   contains  information about the Majorana fermions.

\end{abstract}

\maketitle


\vspace{0.2 in}

\textbf{I-Introduction}

\vspace{0.2 in}

 One of the important ideas  in Condensed Matter Physics is the concept  of topological order \cite{Volkov,Haldane,Golterman,Kreutz,Thouless,Berry,Mele,Kane,davidSpinorbit, More,ZhangField,Hasan,Chao,Zhangnew,Simon,rings,genus,PMC}. Insulators   with a single Dirac cone  which lies in a gap  such  as    $Bi_{2}Se_{3}$, $Bi_{2}Te_{3}$  and $Bi_{1-x}Sb_{x}$ \cite{Volkov,Hasan}  represent the  experimental realization of  Topological Insulators ($T.I$) and recently the Weyl semimetal \cite{Murakami,Vish,Soluyanov,F.Haldane} such as $WTe_{2}$  and  $HgCr_{2}Se_{4}$.

 At the surface   of the three dimensional  Topological Insulator ($T.I.$) one  obtains a two dimensional  metallic surface  characterized  by  an odd number of chiral excitations,  due to  Kramers theorem,   electrons are  protected against backscattering  \cite{balatsky} and localization  \cite{Hai,davidT}. When time reversal is broken  localization effects are observed \cite{Ando}. The surface  physics  has been realized  in     $CdTe/HgTe/CdTe$  quantum wells.   The  quantized  spin-Hall effect has been proposed    \cite{Haldane} and  observed  \cite{Zhangnew,Wu} and recently the Anomalous Hall effect has been measured \cite{Takahashi}.
The  spin resolved photoemission  \cite{Hasan} has been used to identify the surface states.  Topological superconductors  and their identification through the Majorana Fermions have  been observed \cite{Alicea}.

The purpose of this  paper is to  present a geometrical  formulation based on the momentum  representation  of the coordinates   in  periodic solids.  This method has been introduced  in the context of the Spin-Hall effect \cite{davidSpinorbit}.
In order to study  the physical properties of  periodic solids  we use the Brilouine Zone ($B.Z.$). Due to the spin degrees of freedom the wave function is replaced by  a spinor. When we compare spinors at different points in the   $B.Z.$ we observe that the spinors rotate   in  the  $B.Z.$. For this reason we need to introduce a method which  compare the spinors at different points in the $B.Z.$. The parallel  transport \cite{Nakahara} method is best suited   for such problems. 
In    momentum space  the coordinate   $\vec{x}$ is given by the momentum  derivative  $i\vec{\partial}_{k}$. Since the spinors depend on the momentum  the derivative will be covariant  and will introduce a  spin connection. These  comparison of spinors at different points in the $B.Z.$  is done with the help of the   spin connections  (gauge fields)  \cite{davidSpinorbit}.  For   different cases we have either  a  multivalued spin  connections, obstructions or degeneracies \cite{Nakahara}.   The discrete  symmetries:   time reversal symmetry, parity inversion, mirror  symmetry or  charge conjugation    acts as constraints  which fix the  topological  properties.
Due to the symmetries the   eigenvectors   satisfy  certain constraint equations.  The solutions  of  the constraint symmetry gives rise to   specific gauge symmetry for the spin connections in the momentum space.  This gauge symmetry is  used to  compute the electromagnetic response, determine  the   topological invariants .    The electromagnetic  (magnetoelectric) response  for the  $T.I.$ is characterized by  the \textbf{second Chern number}. For the Weyl Semimetal one of the symmetries,  time reversal  or inversion are  broken and as result one has an even number of  Weyl nodes connected by Fermi  arcs.  The  \textbf{ Berry curvature for  the  Weyl  nodes are at  opposite momentum and gives rise  to monopoles and antimonopoles}.

 The interplay of \textbf{Topological Insulators} ($T.I.$), Weyl metals  and \textbf{Superconductivity}  gives  rise to \textbf{Majorana Fermions}

The plan of this review paper is as follows:
 In section $II$ we introduce the method of covariant coordinates in the momentum space. We show that this approach   gives rise to the concepts of spin connections and curvature. In section $III$ we use the method to study the surface properties of a $TI$. We show that the spin connections  for $T^2=-1$  gives rise to anti-localization. Section IV is devoted to the Heisenberg equation of motion given in terms of the covariant coordinates .  Section V  is devoted to the computation of the Topological invariants for the $T.I.$. In section VI we compute the Topological invariants for Topological Crystals.  Section $VII$ is devoted to Weyl Semimetals. In section $VIII$  we investigate a $T.I.$ for  a curved surface and show that the eigenvalues are   effected by the curvature.   Section $IX$ is devoted to  topological invariant for  Superconductors. In   section $X$ we consider the $p-wave$ superconductors which emerge on the  $TI$ surface .   Section $XI$ is devoted to the Andreev crossed reflection and computation of the differential conductivity. In section $XII$ we show that the Majorana fermions  can be detected with a  Squid in a persistent current experiment.   In section $XIII$ we present our conclusions.

\vspace{0.2 in}

\noindent \textbf{ II -The method of parallel  transport in momentum space, spin connection and curvature -essential tools for Topological Insulators}

\vspace{0.2 in}

 Topological Insulators and Superconductors   are gaped  fermionic systems which exhibit topological protected boundary for an  arbitrary deformation, as long the  discrete symmetry such as \textbf{ time reversal, particle hole and chiral symmetry } are respected.
.
 Due to the  symmetries the  Hamiltonian  in the momentum space  $h(\vec{k})$ is invariant at the time  reversal invariant point $\vec{k}=\vec{\Gamma}$  (points in the Brillouin zone) which obey $Th( \vec{\Gamma})T^{-1}=h(-\vec{\Gamma})$ ($T$ is the time reversal operator which obeys  $T^2=-1$) or charge conjugation $Ch( \vec{k})C^{-1}=-h^{*}(-\vec{k})$  ($C^2=-1$). The product of the charge conjugation (particle-hole) with the time reversal allows to define the \textbf{unitary chiral symmetry} which holds in the entire  $B.Z.$. As a result we have $10$   symmetry classes. 
For the case that the inversion symmetry  $P$ and the  time reversal  $T$ symmetry hold, it has been proposed \cite{Kane,Fu} that computing the eigenvalue for  the  ground state parity state,   $\langle U^{(-)}(\vec{\Gamma})|P|U^{(-)}(\vec{\Gamma})\rangle= \mathbf{Sgn}M(\Gamma)$  ($M(\Gamma)$  is the gap at the momentum $\vec{\Gamma}$) determine the topological invariant index $Z_{2}$ .  This index is given by the product   $I=\prod_{i}\mathbf{Sgn}M(\Gamma_{i})$  of the parity eigenvalues at the time reversal invariant point $\vec{\Gamma_{i}}$ in the  $B.Z.$
 The index $I$  determines the coefficient  of   the \textbf{quantized  electromagnetic response} \cite{ZhangField,Zhangnew}.

The  spinors in the  presence of  the  spin-orbit interaction \cite{davidSpinorbit}  vary from point to point.
We will work with the gauged transformed Hamiltonian  $h(\vec{k})$
defined as $e^{i\vec{k}\cdot\vec{x}}h(-i\partial_{x},\vec{x})e^{-i\vec{k}\cdot\vec{x}}= h(\vec{k},i\partial_{k}+\vec{A}(\vec{k}))$.
$\vec{A}(\vec{k})$ is the spin connection which emerges from the covariant  derivatives.  
\noindent 

 For translational invariant systems  
 the Bloch spinor are given  by   $|\vec{k}\otimes \eta_{\alpha}(\vec{k})\rangle \equiv  |U_{\alpha}(\vec{k})\rangle$  (when orbitals are included  $|\vec{k}\otimes \sigma_{i}(\vec{k})\otimes  \tau_{s}( \vec{k})\rangle \equiv |U_{\alpha}(\vec{k})\rangle$ ). As a result we have the representation,
\begin{equation}
|\psi(\vec{k})\rangle= \sum_{\alpha} C^{\alpha}(\vec{k})|U_{\alpha}(\vec{k})\rangle
\label{transformation}
\end{equation}
 When the coordinate acts on the spinor we obtain:
\begin{equation}
x_{a}|\psi(\vec{k})\rangle = i\partial_{k_{a}}|\psi(\vec{k})\rangle=  i\partial_{k_{a}}\sum_{\alpha}C^{\alpha}(\vec{k})|U_{\alpha}(\vec{k})\rangle
\label{ta}
\end {equation} 
$\alpha$ is the index of the band and includes also the spin degree of freedom.
Due to the spin connections the derivative is replaced  by   the covariant derivative:
\begin{eqnarray}
&&|\nabla_{a}\psi(\vec{k})\rangle=\sum_{\alpha}\Big[ \partial_{a}  C_{\alpha}(\vec{k})    -i A^{\alpha,\gamma}_{a}(\vec{k}) C_{\gamma}(\vec{k})  \Big] |U_{\alpha}(\vec{k})\rangle\nonumber\\&&
\hat{A}_{a}(\vec{k})\equiv A^{\alpha,\beta}_{a}(\vec{k})|U_{\alpha}(\vec{k})\rangle \langle U_{\alpha}(\vec{k})|,\hspace{0.1 in}
-i A^{\alpha,\beta}_{a}(\vec{k})\equiv \langle U_{\alpha}(\vec{k})|\partial_{a}|  U_{\beta}(\vec{k})\rangle=\int\,dx U^{*}_{\alpha,k}(x)(-i X^{a}) U_{\beta,k}(x)\nonumber\\&&
X_{a}=x_{a}+A_{a}(\vec{k})\nonumber\\&&  
\end{eqnarray}
$ x_{a}$ is the coordinate in the momentum space,   $ x_{a}=i\partial_{k^{a}}$  which  obeys $[x^a,k_{b}]=[X^{a},k_{b}]= i\delta_{a,b}$. The covariant coordinate is  defined  by $\nabla_{a} \equiv X_{a}=x_{a}+A_{a}(\vec{k})$.

\noindent 
The covariant derivative  introduces the concept of parallel transport, which shows how spinors are translated from point to point. The parallel transport condition is $\langle\psi(\vec{k}) |\nabla_{a}\psi(\vec{k})\rangle=0$ which gives rise to the equations
\begin{equation}
 | \Psi(\vec{k},\hat{P})\rangle= \hat{P} \Big[e^{-i\int_{-\infty}^{\vec{k}}\,dk'^{a} A_{a}(\vec{k}')}\Big]|\psi(\vec{k})\rangle, \hspace{0.1 in}
|\psi(\vec{p})\rangle= = \hat{P}  \Big[e^{-i\int_{\vec{p}}^{\vec{k}}\,dk'^{a} A_{a}(\vec{k}')}|]|\psi(\vec{k})\rangle
\label{transport}
\end{equation}
Where $\hat{P}$ defines an ordered path.  
From the spin connection operator  $\hat{A}_{a}(\vec{k})$ \cite{Nakahara} we construct the curvature operator- $\hat{F}(\vec{k})$.
This is done with the help of the exterior derivative $d$.
Acting with the operator $d$ on the eigenvector $|U_{\alpha}(\vec{k})\rangle$,  $id|U_{\alpha}(\vec{k})\rangle$ we obtain the representation of the  spin connections  $\hat{A}_{a}(\vec{k})\equiv\sum_{\alpha,\beta} A^{\alpha,\beta}_{a}(\vec{k})|U_{\alpha}(\vec{k})\rangle \langle U_{\beta}(\vec{k})|$.
The curvature operator  $\hat{F}_{a,b}(\vec{k})$ is obtained from   the spin connections $A^{\alpha,\beta}_{a}(\vec{k})$.
\begin{eqnarray}
&&\hat{F}_{a,b}(\vec{k})  \equiv [X_{a}(\vec{k}),X_{b}(\vec{k})]= \sum_{\alpha,\beta} F^{(\alpha,\beta)}_{a,b}(\vec{k}) |U_{\alpha}(\vec{k})\rangle \langle U_{\beta}(\vec{k})| \nonumber\\&&
  F^{(\alpha,\beta)}_{a,b}(\vec{k}) =\partial_{a} A^{\alpha,\beta}_{b}(\vec{k})-\partial_{b} A^{\alpha,\beta}_{a}(\vec{k}) +i[\partial_{a} A^{\alpha,\gamma}_{b}(\vec{k}),\partial_{b} A^{\gamma,\beta}_{a}(\vec{k})]\nonumber\\&&
\end{eqnarray}
The curvature allows us  to introduce   the first and second \textbf{chern character}\cite{Nakahara}.
\begin{equation}
ch_{1}= \frac{i}{2\pi}\mathbf{Tr}\Big( F_{a,b}(\vec{k})\Big),\hspace{0.1in} ch_{2}= \frac{-1}{32 \pi^2} \epsilon^{a,b,c,d}\mathbf{Tr}\Big( F_{a,b}(\vec{k}) F_{c,d}(\vec{k})\Big)
\label{curv}
\end{equation}

\noindent In   \textbf{Appendix A}  we   will show  how the curvature  is derived from the eigenfunctions.

\vspace{0.2 in}

 \textbf{III- Observing the topology   using external sources  or disorder potential-transport of Topological Insulators}

\vspace{0.2 in}

\noindent The Hamiltonian can  be express in terms of the eigenvalues.
To obtain   information about the topology, we have to transport the spinor around the B.Z..
Alternatively we  can couple  the  scalar potential $a_{0}(\vec{x})$ to the electronic density  $\rho(\vec{x})$ and the vector potential $\vec{a}(\vec{x})$ to the current density $\vec{J}(\vec{x})$.

\noindent
 For  spin  half  Hamiltonian with  two orbitals 
 $|\vec{k}\otimes \sigma=\uparrow,\downarrow\otimes \tau=1,2\rangle$ for  the  $T.I.$ Hamiltonian  $ H_{0}$ we find:  
\begin{equation}
 H=H_{0}+\int \,d^d x[\rho(\vec{x})a_{0}(\vec{x})+\vec{J}(\vec{x})\cdot \vec{a}(\vec{x})]\equiv H_{0}+H^{ext.} 
\label{ham}
\end{equation}
\noindent The four component spinors for the Hamiltonian $H_{0}$:
\begin{equation}
 \psi(\vec{x})=\int\,d^dk e^{i\vec{k}\cdot\vec{x}} \psi(\vec{k}),\hspace{0.1  in}
 \psi(\vec{k})=\sum_{s=1,2}[ C_{s}(\vec{k})U_{s}(\vec{k})+B^{\dagger}_{s}(-\vec{k})V_{s}(-\vec{k})]
\label{spin}
\end{equation}
 $U_{s}(\vec{k})$  is the four component spinor for the particles  and $ V_{s}(\vec{k})$   is the  four component  spinors for the antiparticles. $s=1,2$ represents the spin helicity.
\begin{eqnarray}
&&H^{ext.} = \int\,d^{d} x \psi^{\dagger}(\vec{x})V(\vec{x})\psi(\vec{x})\nonumber\\&&=\int\, d^{d} k \int\, d^{d} q \sum_{s,s'}\Big[C^{\dagger}_{s}(\vec{k})V(\vec{q})\langle U_{s}(\vec{k})|U_{s'}(\vec{k}+\vec{q})\rangle C_{s'}(\vec{k}+\vec{q})\Big]\nonumber\\&&
\approx\sum_{a} \int\, d^{d} q  V(\vec{q}))(-iq^{a}) \sum_{s,s'}  \int\, d^{d} k C^{\dagger}_{s}(\vec{k}) \Big[\delta_{s,s'}i\partial_{a}+A_{a}^{(s,s')}(\vec{k})\Big]C_{s'}(\vec{k})\nonumber\\&&
i\delta_{s,s'}\partial_{a}+A_{a}^{(s,s')}(\vec{k})\equiv \hat{X}^{a},\hspace{0.1in}
\Big[\hat{X}^{a},\hat{X}^{b}\Big]\equiv \hat{F}_{a,b}(\vec{k})
\end {eqnarray}
 On the  surface  of a $T.I.$ (Topological Insulators) the 
T.R.S. (time reversal invariance) gives  
 for  the  spin-orbit scattering,   the integrated \textbf{ Fermi Surface} curvature is    $\pi$ which gives rise to  $anti-localization$.
This can be   seen from the properties  of  the spin connections. 
   The spin connections  change sign for the transformation  $\vec{k}\rightarrow- \vec{k}$. This change of signs  occurs   also  for the  Cooperon channel!
As a result the conductivity increases.
The  two dimensional $TI$ Hamiltonian  with the scattering potential $V(\vec{x})$ is given by:
\begin{equation}
h(\vec{k} ,\vec{x}) =-\sigma_{2} k_{1}+\sigma_{1} k_{2}+V(\vec{x})
\label{spi}
\end{equation}
For a finite chemical potential $\mu>0$   we consider only the conduction band. We find in terms of the eigenvectors   $U_{\sigma}(\vec{k})$ the representation:
\begin{eqnarray}
&&\Psi_{\sigma}(\vec{k})= C(\vec{k})  U_{\sigma}(\vec{k})\nonumber\\&&
U_{\sigma=\uparrow}(\vec{k})=\frac{1}{\sqrt{2}}e^{\frac{i}{2}\chi(\vec{k})}; U_{\sigma=\downarrow}(\vec{k})=\frac{1}{\sqrt{2}} i e^{\frac{i}{2}\chi(\vec{k})}; 
\chi(-\vec{k})=\chi(\vec{k})+\pi\nonumber\\&&
\end{eqnarray}
We find for the conductions electron the    scattering  potential   gives rise to the  Hamiltonian:
\begin{eqnarray}
&&H^{ext.}=\int\,d^2x \Psi^{\dagger}(\vec{x})V(\vec{x})\Psi(\vec{x})=\sum_{\sigma,\sigma'}
\int\, d^{2} k \int\, d ^{2}q \Big[V(\vec{q})C^{\dagger}(\vec{k})\langle U^{*}_{\sigma}(\vec{k})|U_{\sigma'}(\vec{k}+\vec{q})\rangle C(\vec{k}+\vec{q})\Big]\nonumber\\&&
\approx \sum_{a} \int\, d^2 q  V(\vec{q}))(-iq^{a})  \int\, d ^{2}k C^{\dagger}(\vec{k}) \Big[i\partial_{a}+A_{a}(\vec{k})\Big]C(\vec{k}),\hspace{0.1 in}\partial_{a}
A_{a}(\vec{k})=\partial_{a}\frac{1}{2}\chi(\vec{k})\nonumber\\&&
\end{eqnarray}
 Eq.$(12)$ determines the scattering matrix $S$. We find that the  scattering matrix $S$ obeys:
\begin{equation}
S(\vec{k}\rightarrow -\vec{k})=e^{i\pi}TS(\vec{k}\rightarrow -\vec{k})=-S(-\vec{k}\rightarrow \vec{k})
\label{scattering}
\end{equation}

\vspace{0.2 in}

  \textbf{IV-Equation of motion in the B.Z. for non commuting coordinates -an exact formulation}

\vspace{0.2 in}

In this section we will use the Heisenberg equation of motion  using the representation introduced in the previous chapter:  $ x_{a}=i\partial_{k^{a}}$ , $[x^a,k_{b}]=[X^{a},k_{b}]= i\delta_{a,b}$, $ X_{a}=x_{a}+A_{a}(\vec{k})$.
In addition, for a non uniform system we will have a real space curvature similar  to the magnetic field.

The methodology of a "curved" space induced by  the spin orbit coupling in the B.Z. was  described  in  terms  of the spin  connections  $ A^{\alpha,\beta}_{a}(\vec{k})\equiv  \langle U_{\alpha}(\vec{k})|i\partial_{a}|  U_{\beta}(\vec{k})\rangle$ and the  \textbf{curvature} $[X_{a},X_{b}]=F_{a,b}$ . 
 
\noindent 
 The Hamiltonian in the  eigenvalue representation  is given by: $\mathbf{h}(\vec{k})=\sum_{\lambda}E_{\lambda}(\vec{k})|U_{\lambda}(\vec{k})><U_{\lambda}(\vec{k})|$  the presence of a scalar potential $\mathcal{V}(\vec{r})$ is replaced by $\mathbf{V}(\vec{R})$  where  $\vec{R}$ is the  covariant coordinate. We find for the  Heisenberg equation :
\begin{eqnarray}
&&\hat {h}(\vec{k},\vec{R})\equiv\mathbf{h}(\vec{k})+\mathbf{V}(\vec{R})\nonumber\\&&
\frac{d k^{a}}{dt}=\frac{-1}{\hbar}\partial_{R_{a}}\mathbf{V}(\vec{R}) \nonumber\\&& \frac{d R_{a}}{dt}=\frac{1}{\hbar}\partial_{k^{a}} \mathbf{h} (\vec{k})+\frac{1}{\hbar}\epsilon^{a,b,c}F_{b,c}\partial_{R_{a}}\mathbf{V}(\vec{R})\nonumber\\&&
\end{eqnarray}
When the real space is curved, due to dislocations or magnetic fields the momentum operator $k^{a}$ is replaced by a covariant momentum $K^{a}=k^{a}-\frac{i}{4}Log[g]$ where  $g=\sqrt{ Det g_{\vec{r},t}}$ is the metric tensor  \cite{davidSpinorbit}. For this situation we introduce  the real  space momentum  curvature  $ \Omega_{a,b}=[K^{a},K^{b}]$ which together with coordinates curvature  $F_{a,b}=[X_{a},X_{b}] $ modifies the equations of motion.
\begin{eqnarray}
&&\frac{d K^{a}}{dt}=\frac{-1}{\hbar}\partial_{R_{a}}\mathbf{V}(\vec{R}) +\frac{1}{\hbar}\epsilon^{a,b,c}\Omega_{b,c}\partial_{K^{a}}\mathbf{h}(\vec{K})\nonumber\\&&  \frac{d R_{a}}{dt}=\frac{1}{\hbar}\partial_{K^{a} }\mathbf{ h}(\vec{K})+\frac{1}{\hbar}\epsilon^{a,b,c}F_{b,c}\partial_{R_{a}}\mathbf{V}(\vec{R}) \nonumber\\&&
\end{eqnarray} 

\vspace{0.2 in}

\vspace{0.2 in}

\textbf{V-Topological invariants from response theory}

\vspace{0.2 in}

\noindent We will  demonstrate how  the topological invariant  appears.
 In this chapter we will  appreciate the power of  the covariant coordinates and their commutations. We observe that the topological invariants are given by the commutators of the covariant coordinates. 

We  will consider a typical  $T.I.$   Hamiltonian  for the materials  $ Bi_{2}Se_{3}$, $ Bi_{2}Te_{3}$, $Sb_{2}Te_{3}$ . We  introduce    the  tensor product  $|\alpha\rangle\equiv|\sigma=\uparrow,\downarrow\rangle\otimes |\tau=1,2\rangle$    ($\sigma$ stands for the $ spin$  and    $\tau$  stands for the  orbitals).   A four band model   is obtained \cite{Chao}  which  can be  written in  the chiral form  \cite{Ramond} :
$h(\vec{k})=I \epsilon(\vec{k})+\gamma_{1}\hat{k}_{2}+\gamma_{2}\hat{k}_{1}+ \gamma_{3}\eta\hat{k}_{3}+ \gamma_{0}M(\vec{k})$.

\noindent  The first term affects only the eigenvalues  and not the eigenvectors. The parameter  $\eta$ satisfies  $\eta<<1$.
 The $ \gamma$  matrices  are given as a tensor product  :$\gamma_{i}=\sigma^{i}\otimes(-1) \tau_{2}$, $i=1,2,3$ , $\gamma_{0}=I\otimes\tau_{1}$ ,$\Gamma_{5}=I\times \tau_{3}$.
 The mass  (gap)  $M(\vec{k})$ obeys $M(-\vec{k})=M(\vec{k})$ and has  points in the Brillouin  Zone where it vanishes. (On a lattice with the lattice constant $a$ we define     the Cartesian component of the  momentum  $\hat{k}_{i}=\frac{ \sin[k_{i} a]}{a}$.) 
 The Hamiltonian $h(\vec{k})$ is  diagonalized using the four eigenvectors $ |U_{s}^{(e)}(\vec{k})\rangle$, $s=1,-1$ are  the  \textbf{spin helicity} operator and $e=+,-$  represents  the particles-antiparticles energies,  $E(\vec{k})= \epsilon(\vec{k}) \pm\sqrt{k^2+M^{2}(\vec{k})}$ with the  mass   $M(\vec{k})$ which vanishes at $\vec{k}^{*}$ and   $M(\vec{\Gamma})\neq0$.The Hamiltonian in the eigenvalue representation is given by,
\begin{equation}
 h(\vec{k})=\sum_{s=1,-1}\Big[E(\vec{k})|U_{s}^{(e=+)}(\vec{k})\rangle \langle U_{s}^{(e=+)}(\vec{k})| + E(\vec{k})|U_{s}^{(e=+)}(\vec{k})\rangle\langle U_{s}^{(e=+)}(\vec{k})|\Big]
\label{bas}
\end{equation}
\noindent    The Green's function operator in the  $\alpha$  basis $|\alpha\rangle\equiv |\sigma=\uparrow,\downarrow\rangle\otimes|\tau=1,2\rangle$  is given by:
 $\hat{G}(\omega,\vec{k})=\sum_{\alpha,\alpha'} \hat{G}(\omega,\vec{k})_{\alpha,\alpha'}|\alpha\rangle \langle \alpha'|$. 

In the \textbf{eigenvector   basis}  the Green's  function takes the form:
\begin{equation}
\tilde{G}(\omega,\vec{k})=\sum_{s=1,-1}[\frac{ |U_{s}^{(e=+)}(\vec{k})\rangle\langle U_{s}^{(e=+)}(\vec{k})|}{ \omega - E(\vec{k})+i\epsilon}+ \frac{ |U_{s}^{(e=-)}(\vec{k})\rangle \langle U_{s}^{(e=-)}(\vec{k})|}{ \omega + E(-\vec{k})+i\epsilon}]
\label{equation}
\end{equation}

\noindent
 The transformation from  the  $|\alpha\rangle$   basis to the eigenvector  basis  $|U_{s}^{e}(\vec{k})\rangle$  replaces the coordinate $\hat{r}_{i}=i\partial_{k}^{i}$  with  \textbf{ the covariant coordinate $\hat{R}_{i}$} \cite{davidSpinorbit}.
\noindent  In the second quantized form  the spinor operator $\Psi(\vec{r})$ is given by: 
\begin{equation}
\Psi(\vec{r})=\int\,\frac{d^d k}{(2\pi)^{d}}  e^{i\vec{k}\cdot \vec{r}}\Psi(\vec{k}),\hspace{0.02 in}
\Psi(\vec{k})=\sum_{s=1,-1}\Big[C_{s}(\vec{k})U^{(+)} _{s}(\vec{k}) +  b^{\dagger}_{s}(-\vec{k})U^{(-)} _{s}(-\vec{k})\Big],\hspace{0.01 in} 
 \bar{\Psi}(\vec{k})=\Psi^{\dagger}(\vec{k})\gamma_{0}
\label{eqw}
\end{equation}
\noindent
 (It is important to stress that this representation is valid for momentum $|\vec{k}|< |\vec{k}^{*}|$, for  the region $|\vec{k}|> |\vec{k}^{*}|$  we need to choose a different representation.)The coupling of the $T.I.$  to the electromagnetic field  $\vec{a}(\vec{r},t)$ and  $ a_{0}(\vec{r},t)$ is given by the action   $S^{ext}$:
\begin{eqnarray} 
&&S^{ext}=\int\,\frac{d^d k}{(2\pi)^{d}}\int\,\frac{d^d Q}{(2\pi)^{d}} \int\,\frac{d \omega}{2\pi} \int\,\frac{d \Omega}{2\pi}[  \bar{\Psi}(\vec{k},\omega) \gamma_{\nu}a_{\nu}(\vec{Q},\Omega)\Psi(\vec{k}+\vec{Q},\omega+\Omega)] \nonumber\\&&  \approx\int\,d^3 x\int\,dt[\sum_{\nu=0}^{3}\sum_{\mu=0}^{3}\gamma_{\nu}(\partial_{\mu}a_{\nu}(\vec{r},t))|_{\vec{r}=0,t=0)} ]\int\,\frac{d^{d}k}{(2\pi)^d}\int\,\frac{d\omega}{2\pi}
\bar{\Psi}(\vec{k},\omega)\gamma^{\mu}\hat{R}_{\mu}\Psi(\vec{k},\omega)]\nonumber\\&&
\bar{\Psi}(\vec{k},\omega)=\Psi^{\dagger}(\vec{k},\omega)\gamma_{0};  \hat{R}_{\mu=0}=i\partial_{\omega}\nonumber\\&&
\end{eqnarray}  

\noindent
We compute the  \textbf{partition function}  \cite{Ramond} for four  space dimensions. We  find that  the effective action for  the electromagnetic fields \cite{Golterman} is given by $\Gamma[a_{0}(t,\vec{r}),\vec{a}(t,\vec{r})]$:
\begin{equation}
 Z=\int\prod_{s}\prod_{k}\prod_{\omega}dC_{s}(\vec{k},\omega)d C^{\dagger}_{s}(\vec{k},\omega)db_{s}(-\vec{k},\omega)d b^{\dagger}_{s}(-\vec{k},\omega) e^{i(  S^{0}+S^{ext})}=e^{i \Gamma[a_{0}(t,\vec{r}),\vec{a}(t,\vec{r})]}
\label{action}
\end{equation}

\noindent Using the totally antisymmetric tensor $\epsilon_{\alpha_{ 1},\alpha _{2},\alpha_ {3},\alpha _{4},\alpha_{5}}$ for  four space dimensions  we find    the electromagnetic  response which is given in terms of the electric $\vec{E}$ and the magnetic  field $\vec{B}$.

\noindent   
The  coefficient $\hat{C}_{2}$ is  quantized, for values of  $\theta=\pi,0$  when the time reversal symmetry  is preserved \cite{Golterman}.

\noindent 
The electromagnetic  response   in     $4+1$ space time  dimensions is \cite{Golterman,Weinberg,Nakahara,davidtop,ZhangField,Zhangnew}:
\begin{equation}  
\hat{C}_{2} =const.\epsilon^{0,i,j,k,l}Tr[\tilde{G}(\omega,\vec{k})\hat{R}_{0}\tilde{G}^{-1}(\vec{k})\tilde{G}(\omega,\vec{k})\hat{R}_{i}\tilde{G}^{-1}(\omega,\vec{k})\tilde{G}(\omega,\vec{k})\hat{R}_{k}\tilde{G}^{-1}(\omega,\vec{k})\tilde{G}(\omega,\vec{k})\hat{R}_{l}\tilde{G}^{-1}(\omega,\vec{k})]
\label{iint}
\end{equation}

 In the presence of  electron-electron interactions  the \textbf{Green's function  can    have   zero's} \cite{Gourarie,Zhong},  for such a case the system is not  topological! If the renormalized Green's  function has \textbf{no zero's the topology is preserved}. The renormalized Green's function   $\tilde{G}_{R}(\omega,\vec{k})=\sum_{s}\frac{Z^{-1} (\frac{\kappa}{\Lambda},u_{i},M)}{(\omega+i\epsilon-E_{s}(\vec{k}))}|U_{s}(\vec{k})\rangle \langle U_{s}(\vec{k})|$ is given in terms  of the wave function renormalization  $Z (\frac{\kappa}{\Lambda},u_{i},M)=\Big[1-\frac{\partial_{\omega}\Sigma(\frac{\omega}{\Lambda},u_{i},M)}{\partial_{\omega}}|_{\omega=\kappa}\Big]^{-1}$. 

\noindent
 When the wave function renormalization $Z$ is finite at  $\omega=0$  we take the limit  $\omega\rightarrow 0$ and obtain the second Chern  character  \cite{Zhong}, $Z$   cancels:
\begin{equation}
 \hat{C}_{2} =const.\epsilon^{i,j,k,l}Tr[\tilde{G}(\vec{k})\hat{R}_{i}\tilde{G}^{-1}(\vec{k})\tilde{G}(\vec{k})\hat{R}_{j}\tilde{G}^{-1}(\vec{k})\tilde{G}(\vec{k})\hat{R}_{k}\tilde{G}^{-1}(\vec{k})\tilde{G}(\vec{k})\hat{R}_{l}\tilde{G}^{-1}(\vec{k})]
\label{cc}
\end{equation}
The trace operator acts only on the  occupied bands.   The Green's function in the eigenvector basis  representation  makes use  of     the  \textbf{covariant}  matrix coordinates  $\hat{R}_{i}$. The Chern character   is given as  a matrix  product.
 The   commutator   $[ \hat{R}_{i},\hat{R}_{j}]=F_{i,j}(\vec{k})$
gives the curvature   $ F_{i,j}(\vec{k})$   in terms of  the  spin connections
$ \mathcal{A}^{(s,s')}_{i}(\vec{k})$. 
\noindent   The second Chern number is given by,
\begin{equation}
  C_{2}=\frac{1}{32\pi^2}\int\,d^4k\epsilon^{i,j,k,l}Tr[ F_{i,j}(\vec{k})F_{k,l}(\vec{k})]
\label{top}
\end{equation}
  The integral in Eq.$(23)$  is either  zero (exact form) or non-zero  (non exact form).  Therefore,
\begin{equation}
 Tr[ F_{i,j}(\vec{k})F_{k,l}(\vec{k})]\equiv Tr[F^2] = d[K_{3}] 
\label{exact}
\end{equation}
where $K_{3}=Tr[AdA+\frac{2}{3}A^3]$    is the \textbf{Chern-Simons}  three form  \cite{Nakahara} .
 $K_{3}$  can be found with the help of  the  gauge symmetry imposed by  $T^2=-1$ \cite{Nakahara}. 
 The second Chern number in four dimensional space is  given by: $C_{2}=\frac{1}{32\pi^2}\int_{BZ}\,d^{4} k\epsilon^{i,j,k,l} Tr[F_{i,j} F_{k,l}]$  which has a $Z_{2}$  winding number.
 $Tr[F^2]$ is   $closed$ but not $exact$ \cite{Nakahara}.  This means that in some restricted regions of  the Brillouin zone the integral $\int_{S^{4}}Tr[Tr[F^2]$  is  given by  a  Chern-Simons contour integral \cite{Nakahara}:
\begin{equation}
Tr[{F}^2]= d[K_{3}]=d(Tr[\mathcal{A}d\mathcal{A}+\frac{2}{3}\mathcal{A}^3])
\label{chern}
\end{equation}
 This result does not hold  in the entire $B.Z.$.  We can identify two regions which are related by a transition  function (a gauge transformation), a transformation matrix between states  for  the  region $\vec{k}$ which belong to half of the positive sphere    $(S^{4})_{+}$ and  $\vec{k}$ which belong to second half of the negative sphere   $(S^{4})_{-}$ .

\noindent 
Next we compute the effect of the transformation $\vec{k}\rightarrow -\vec{k}$ on the spin connections. Due to the time reversal symmetry 
we find   the matrix which transforms between the two regions  of $\vec{k}$ . The transformation is given by the \textbf{Pfaffian} matrix  $B$ defined in terms of the Kramers pair. 
 We  use the eigenvectors  to compute the  matrix   $B_{ s,s'}(\vec{k})$ ($T=-i\sigma_{2}K$, $K$  is the  complex conjugation operator),
\begin{equation}
 B_{ s,s'}(\vec{k}) =\langle U_{s}^{(-)}(-\vec{k})|T |U_{s'}^{(-)}(\vec{k})\rangle
\label{matrix}
\end{equation}
 The relation between the spin connections  $\mathcal{A}(\vec{k})$  and $\mathcal{A}(-\vec{k})$ is given by:
\begin{equation}
\mathcal{A}(-\vec{k})= B(\vec{k})\mathcal{A}^{*}(\vec{k})B^{\dagger}(\vec{k}) +i  B(\vec{k})\partial_{k^{i}}B^{\dagger}(\vec{k})
\label{connection}
\end{equation}
As a result  the  curvatures transform like :
\begin{equation}
 F_{i,j}(-\vec{k})=-B(-\vec{k})  F^*_{i,j}(\vec{k})B(-\vec{k})
\label{curvature}
\end{equation}
Applying Stokes theorem  for $ d[K_{3}]$ in  the two  regions one  obtains a boundary integral over the  difference  of the  Chern-Simons terms (defined for each region). The difference between the two Chern-Simons terms  can be understood as a polarization \textbf{$P(\vec{q})$}   difference  between   two regions. (The boundary term  is a surface perpendicular to   the fourth direction $\vec{q}$. )
\begin{eqnarray}
&& C_{2}=\frac{1}{32\pi^2}\int_{BZ}\,d^{4} k\epsilon^{i,j,k,l} Tr[F_{i,j} F_{k,l}]=\int_{  \partial S^{4}}([Tr[\mathcal{A}_{+}d\mathcal{A}_{+}+\frac{2}{3}\mathcal{A}_{+}^3]-Tr[\mathcal{A}_{-}d\mathcal{A}_{-}+\frac{2}{3}\mathcal{A}_{-}^3])\nonumber\\&&
= \frac{1}{24\pi^2}\int\,d^{3}k \epsilon_{i,j,k}Tr[(B(\vec{k})\partial_{i}B^{\dagger}(\vec{k}))(B(\vec{k})\partial_{j}B^{\dagger}(\vec{k}))(B(\vec{k})\partial_{k}B^{\dagger}(\vec{k}))] = \sum_{k^*}N_{k^*}=2P[q]\nonumber\\&&
\end{eqnarray} 
 \noindent We have $2P[q]=\sum_{k^*}N_{k^*}$   ($k^{*}$ are points in the Brillouin Zone  where the Pfaffian matrix vanishes) . Due to the lattice periodicity, Bloch theorem  allows  to define   the polarization  \textbf{$P(\vec{q})$}  as  modulo an integer.  We  recover the result $P=0$ for $\sum_{k^*}N_{k^*}=even$ and  $P=\frac{1}{2}$ for  $\sum_{k^*}N_{k^*}=odd$  \cite{Kane}. To conclude  $P$  represents the topological invariant for $d=3$ space obtained from  response theory.

\vspace{0.2 in}

\textbf{VI-Topological invariants   for Topological Crystals}

\vspace{0.2 in}

   The  method used in the previous section  is    applicable to  Topological Crystals where  a mirror  reflection invariant $\eta=(\frac{-i}{\sqrt{2}}(\sigma_{1}+ \sigma_{2})K)$ with the property $\eta^{2}=-1$ replaces the $T^2=-1$ invariant.

\noindent Inspired by \cite{Fu} the authors in ref. \cite{Bansil} proposed that $SnTe$ has a  mirror plane perpendicular to the $[110]$ direction.  A band inversion at  the  four  $L$ points  in the Brillouin  zone between $SnTe$ and $PbTe$  can be achieved for the mixed crystal $Pb_{1-x}Sn_{x}Te$. 
The $k\cdot p$ model near an $L$ point \cite{Bansil} gives the Hamiltonian:
\begin{equation}
h^{L}=(\sigma_{2}\otimes\tau_{1})k_{1}-(\sigma_{1}\otimes\tau_{1})k_{2}+ (I\otimes\tau_{2})+M(\vec{k})(I\otimes\tau_{3}).
\label{mir}
\end{equation}
where $ M(\vec{k})$ is the inverted mass (has zeros in the Brillouin zone), $\sigma=\pm\frac{1}{2}$ corresponds to the states  with total angular momentum $J=\frac{1}{2}$ , $|j=\frac{1}{2},j_{z}=\pm\frac{1}{2}> $  and $\tau =1,2$  corresponds to the $p$ orbitals  of the cation (Sn or Pb) and anion $Te$.

\noindent The mirror invariant $\eta$  with the property   $\eta^2=-1$ can be found  such that  the condition for the polarization is different from the one given  by the time reversal invariant points.

\noindent 
The reflection symmetry from the plane perpendicular to $[1,1,0]$ is given by the transformation $[k_{1},k_{2},k_{3}]\rightarrow [-k_{2},-k_{1},k_{3}]$.  The operator of reflection which acts on the states is given by a rotation of an angle $\pi$ around the axes $[1,1,0]$ and  is accompanied by an inversion trough the origin. 
A simple calculation shows that the mirror operator is given by $M=(\frac{-i}{\sqrt{2}})[(\sigma_{1}+\sigma_{2})\otimes I]$.  As a result the state   $|\phi(\vec{k})>$ is transformed  to $|\phi_{M}(\vec{k}')>$ and the Hamiltonian $h^{L}(\vec{k})$ obeys the symmetry:
\begin{eqnarray}
&&|\phi_{M}(-k_{2},-k_{1},k_{3})>=M|\phi (k_{1},k_{2},k_{3})>\equiv (\frac{-i}{\sqrt{2}})[(\sigma_{1}+\sigma_{2})\otimes I]|\phi (k_{1},k_{2},k_{3})>\nonumber\\&& 
M^{-1}h^{L}(k_{1},k_{2},k_{3})M=h^{L}(-k_{2},-k_{1},k_{3}); \hspace{0.2 in}  M=(\frac{-i}{\sqrt{2}}[(\sigma_{1}+\sigma_{2})\otimes I].\nonumber\\&&
\end {eqnarray}
Next we include the anti unitary  conjugation operator K and define the operator $\eta=MK=(\frac{-i}{\sqrt{2}})[(\sigma_{1}+\sigma_{2})\otimes I]K$ which obeys $ \eta^2=-1$  (this is similar to  the time reversal operator $T=-i\sigma_{2}K$).
We obtain the invariance transformation:
\begin{equation}
\eta^{-1}h^{L}(k_{1},k_{2},k_{3})\eta=(h^{L}(-k_{2},-k_{1},k_{3}))^* .
\label{eta}
\end{equation}
At the     invariance  points  $\vec{k}^{M}$ one obtains the conditions:

 $ h^{L}(-k^{M}_{1},-k^{M}_{2},-k^{M}_{3})= h^{L}(-k^{M}_{2},-k^{M}_{1},k^{M}_{3})$.

For two dimensions we have only two invariant points
 $[0,0]$
and $[\pi,\pi]$ and for
 three dimensions we have   four  invariant points $[0,0,0]$, $[0,0,\pi]$, 
$[\pi,\pi,0]$, $[\pi,\pi,\pi]$.  

At this stage we will   follow the strategy proposed in section  $A$. We  identify the pairs of degenerate eigenvalues.
For this case we expect to find a Kramers  pair for the eigenfunctions $|V_{s=1}(\vec{k};M(\vec{k}))>$ and 
 $|V_{s=-1}(\vec{k};M(\vec{k}))\rangle$.  When the mass parameter $M(\vec{k})$ has zeros   we observe that the matrix $\hat{W}_{s,s'}(\vec{k})$ defined by
 $\hat{W}_{s,s'}(\vec{k})\equiv\langle V_{s}(-\vec{k};M(-\vec{k}))|\eta|V_{s}(\vec{k};M(\vec{k}))\rangle$ 
 is a \textbf{Pfaffian} which vanishes at some $\vec{k}$ ,  $M(\vec{k})=0$. Due to the zero's of the Pfaffian it is not possible to construct a single eigenfunction for the entire Brillouin zone.  We observe  that the eigenfunction at $-\vec{k}$ is related to the eigenfunction with $ \vec{k}$ in the following way: 
\begin{eqnarray}
&&|V _{s=1}(-\vec{k};M(-\vec{k})>=\hat{W}^*_{s=1,s'=-1}(\vec{k})|V _{s'=-1}(\vec{k};M(\vec{k})>\nonumber\\&& |V _{s=-1}(-\vec{k};M(\vec{k})>=\hat{W}^*_{s=-1,s'=1}(\vec{k})|V _{s'=1}(\vec{k};M(\vec{k})>\nonumber\\&&
\end{eqnarray}
 where $M(\vec{k})=M(-\vec{k})$.
 For each pair of mirror invariant  states at the   points $[0,0]$ and $[\pi,\pi]$  (for d=2) and  $[0,0,0]$, $[0,0,\pi]$, $[\pi,\pi,0]$, $[\pi,\pi,\pi]$ (for d=3).    The \textbf{Pfaffian} matrix $W$ induces a transformation for the spin  connections,
resulting in the condition for the  Chern-Simons field  polarization
 $2P[q]=\sum_{k^{M}}N_{k^{M}}$ (with the sum restricted 
to the $\eta$ invariant points).
It results in a different condition for the  polarization,   since $\sum_{k^{M}}N_{k^{M}}$  is different from  the condition $ \sum_{k^{*}}N_{k^{*}}$ for the time reversal case $T=-i\sigma_{2}K$. Therefore we can have a situation where  the polarization is zero according to the time reversal symmetry and non-zero according to the mirror symmetry.

\vspace{0.2 in}

\textbf{VII- The Weyl Semimetal}

\vspace{0.2 in}

In this section we will consider Hamiltonians of the form $H=v_{F}\vec{\sigma} \cdot \vec{b}(\vec{k})$ where $\vec{b}(\vec{k})$ is a vector-valued function $\vec{b}(\vec{k})$ has  singular points $\vec{k}_{r}$ where $\vec{b}(\vec{k})$ vanishes.
Acording  to the Nielson Ninomiya theorem the    singular points $\vec{k}_{r}$ are even and satisfies    $\pm\vec{k}_{r}$. We define the Hamiltonian around these points:
$\sum_{r=1...n}H(\vec{k}=\vec{k}_{r}+\vec{k},\vec{p}=-\vec{k}_{r}+\vec{p})=
\sum_{r=1...n}\pm v^{(r)}_{F}\vec{\sigma} \cdot \vec{p}$.
Around each singularity  we have a monopole or an antimonopole.  The sum of  the monopoles and atimonopoles must be zero in a crystal.

Next we will consider explicit examples:

\noindent
The Dirac metal ($DM$) has two  
 Weyl nodes at a single singularity: 
$H=\pm v_{F}\vec{\sigma} \cdot \vec{k}$ .  When we dope the    $DM$ becomes Dirac semimetal   ($DSM$).

\noindent
A  Weyl Semimetal is characterized by  an even number of  nodes $\pm \vec{k}_{r}$ with an even of point singularity. The   Hamiltonian  has    two opposite  chiralities :
  $H(\pm \vec{k}_{r}+\vec{p})=\pm v_{F}\vec{\sigma} \cdot \vec{p}-\mu$.
where $\mu $ is  the chemical potential.
For each chirality we can compute the eigenspinors $|U_{\alpha}(\vec{k})\rangle$  with the ground state energy $\alpha |k|$.
 For each  chirality  $\alpha=\pm$  we compute the spin  connection, curvature and the divergent  of the curvature  :
\begin{eqnarray}
&&-i A^{\alpha}_{a}(\vec{k})\equiv \langle U_{\alpha}(\vec{k})|\partial_{a}|  U_{\alpha}(\vec{k})\rangle
,\hspace{0.1 in} X^{\alpha}_{a}=x_{a}+A^{\alpha}_{a}(\vec{k})\nonumber\\&&
F^{\alpha}_{a,b}(\vec{k})=\partial_{a} A^{\alpha}_{b}(\vec{k})-\partial_{b} A^{\alpha}_{a}(\vec{k})\nonumber\\&&
\epsilon_{a,b,c}\partial_{a}F^{\alpha}_{b,c}(\vec{k})=\alpha 2\pi\delta(\vec{k}-\alpha \vec{k}_{r}); \hspace{0.1 in} \alpha = \pm\nonumber\\&&
\end{eqnarray}
 For  the  monopole and antimonopole at 
$\pm \vec{k}_{r}$.

\noindent
 The time reversal symmetry and the  inversion symmetry gives  a restriction  for  these curvature and  and the monopoles density. Either inversion symmetry or time reversal can be broken but not both:
For broken inversion we have for the curvature, $F_{a,b}(\vec{k})=F_{a,b}(-\vec{k})$.
For broken time reversal  the curvature obeys, $F_{a,b}(\vec{k})=-F_{a,b}(-\vec{k})$.

\noindent
 On the surface.
of a crystal the  two Weyl nodes are connected by Fermi arcs, for this we have to demand  that in the normal direction of  the surface the wave function is zero. 

\vspace{0.1 in}

\textbf{a) Realization of the Weyl Hamiltonian  for broken inversion symmetry}

\vspace{0.1 in}

We introduce the Pauli matrix  $\vec{\tau}$  for the  two orbitals and $\vec{\sigma}$  for the two spins:
The model which  breaks inversion is given by:

\noindent
$h(\vec{k})=v_{F}\tau_{3}\otimes \vec{\sigma}\cdot \vec{k}+\tau_{1 }\otimes I M(\vec{k})+ \tau_{3 }\otimes I g(\vec{k})$

\noindent
  $M(\vec{k})=M(-\vec{k})$  and $g(\vec{k})= g(-\vec{k})$ are even functions.
  The spinor are given by:
 $\Psi(\vec{k})=\Big[\psi_{1,\uparrow},\psi_{1,\downarrow},\psi_{2,\uparrow},\psi_{2,\downarrow}\Big]^{T}$.

 \noindent 
We  compute  the eigenvalues and find the points in the B.Z. where  $E=0$.



\noindent
According to general theory the low energy Hamiltonian  has at least two  Weyl nodes  
$H=
\pm v^{(r)}_{F}\vec{\sigma}\cdot \vec{p}$.
with the  connections
$A_{a}(\vec{k};\vec{k}_{r})$, $A_{a}(\vec{k};-\vec{k}_{r}))$ and monopole charge  :
 $\epsilon_{a,b,c}\partial_{a}F_{b,c}(\vec{k})=\pm  2\pi\delta\Big(\vec{k}-(\pm \vec{ k}_{ r})\Big)$.

\vspace{0.1 in}

\textbf{b) Realization of the Weyl Hamiltonian  for  broken time reversal  symmetry}

\vspace{0.1 in}

\noindent
Superconductivity with broken time reversal symmetry  has been considered in the literature by  \cite{Murakami,Haldane}.
Here we propose a continuum two band model  with broken time reversal symmetry ($T.R.S.$) given  by:
\begin{equation}
h^{(I)}(\vec{k})=v_{F}(\tau_{1}\otimes \sigma_{3})(k^2_{1}-M)+v_{F}(\tau_{2} \otimes \sigma_{3})k_{2}+v_{F}(\tau_{3} \otimes \sigma_{3})k_{3}+(I\otimes \sigma_{3})B, M>0
\label {equation}
\end{equation}
The  model has  monopoles since the inversion symmetry is not broken.
This model assumes that in a magnetic field $B$ the two bands  are polarized and   the coupling is  linear in the momentum  space (spin orbit interactions and uniaxial stress can realize such a model ). The Pauli matrix $\vec{\tau}$ acts on the $orbital$ space and the  Pauli matrix $\vec{\sigma}$ acts on the spin . 
We will  project the Hamiltonian on the polarized state   $|U_{\uparrow}(\vec{k})\rangle$ and find:
\begin{equation}
h^{(Project)}(\vec{k})=v_{F}\tau_{1}(k^2_{1}-M)+v_{F}\tau_{2}k_{2}+v_{F}\tau_{3}k_{3}-I|B|
\label{equation}
\end{equation}
In order to study  superconductivity we  consider an attractive    short range pairing interactions $H^{paring}$:
\begin{eqnarray}
&&H^{(paring)}=-V(|d|)\int\,d^3x\Big[\Psi^{\dagger}(\vec{x}-\frac{\vec{d}}{2})\Psi^{\dagger}(\vec{x}+\frac{\vec{d}}{2})\Psi(\vec{x}+\frac{\vec{d}}{2})
\Psi(\vec{x}-\frac{\vec{d}}{2})+\Psi^{\dagger}(\vec{x}+\frac{\vec{d}}{2})\Psi^{\dagger}(\vec{x}-\frac{\vec{d}}{2})\Psi(\vec{x}-\frac{\vec{d}}{2})
\Psi(\vec{x}+\frac{\vec{d}}{2})\Big]\nonumber\\&&
\end{eqnarray}
 where $\Psi(\vec{x})=\Big[\psi_{1}(\vec{x}), \psi_{2}(\vec{x})\Big]^{T}$ are the spinor buid from the orbitals $1$ and $2$.
 The paring order parameter must be odd to ensure the inversion  symmetry. We use the Hubbard Stratonovici transformation and rewrite  Eq.$32)$  in momentum space  for a fixed pairing distance $d$.
\begin{equation}
H_{I}=\int\,d^3 k\Big[\frac{|\Delta(d)|^2}{2|V(d)|} +\Big(\psi^{\dagger}_{1}(\vec{k})\psi^{\dagger}_{1}(-\vec{k})+\psi^{\dagger}_{2}(\vec{k})\psi^{\dagger}_{2}(-\vec{k})\Big)i\Delta(d)\sin(\vec{k}\cdot \vec{d})+h.c.\Big]
\label{pairing}
\end{equation}
We linearize  the  spinor around the monopole and anti-monopoles. $\vec{q}=\Big[\pm\sqrt{M},k_{2}=0,k_{3}=0\Big]$ .
As a result  we find the chiral Hamiltonian:
\begin{eqnarray}
&&h_{L}\Big(\vec{k}=[\sqrt{M}+p_{1},p_{2},p_{3}\Big)=2\sqrt{M}\tau_{1}p_{1}-\tau_{2}p_{2}-\tau_{3}p_{3}
\nonumber\\&&
h_{R}\Big(\vec{k}=[-\sqrt{M}+p_{1},p_{2},p_{3}\Big)=-2\sqrt{M}\tau_{1}p_{1}-\tau_{2}p_{2}-\tau_{3}p_{3}
\nonumber\\&&
\end{eqnarray}
Next we expand the spinor in terms of the chiral fermions $C_{L}(\vec{p})$ and $C_{R}(\vec{p})$ determined by  $h_{L}(\vec{k})$ and  $h_{R}(\vec{k})$.
As a result the two component spinor  $\Psi(\vec{p})$ is represented  in terms of the chiral fermions .
$\Psi(\vec{p})=C_{L}(\vec{p})|V(\vec{p}\rangle +C_{R}(\vec{p})|U(\vec{p}\rangle$ . $|V(\vec{p}\rangle$ is a two component eigenvector which represents the two orbitals with left chirality,  similar $|U(\vec{p}\rangle$  has two  components for the two orbitals with the right chirality. 
Using the eigenvectors $|V(\vec{p}\rangle$ ,$|U(\vec{p}\rangle$    we  project the pairing field in terms of the  two chiralities : 
\begin{eqnarray}
&&\psi^{\dagger}_{1}(\vec{p})\psi^{\dagger}_{1}(-\vec{p})+\psi^{\dagger}_{2}(\vec{p})\psi^{\dagger}_{2}(-\vec{p})\approx
C^{\dagger}_{L}(\vec{p})C^{\dagger}_{R}(-\vec{p})\Big(\langle 1| V(\vec{p})\rangle ^*\langle 1| U(-\vec{p}\rangle ^*+\langle 2| V(\vec{p}\rangle ^*(\langle 2| U(-\vec{p}\rangle ^*\Big)\nonumber\\&&
+C^{\dagger}_{R}(\vec{p})C^{\dagger}_{L}(-\vec{p})\Big(\langle 1| U(\vec{p})\rangle ^*\langle 1| V(-\vec{p}\rangle ^*+\langle 2| U(\vec{p}\rangle ^*\langle 2| V(-\vec{p}\rangle ^*\Big)\nonumber\\&&
=C^{\dagger}_{L}(\vec{p})C^{\dagger}_{R}(-\vec{p})F(\vec{p})-C^{\dagger}_{R}(\vec{p})C^{\dagger}_{L}(-\vec{p})F(-\vec{p});
F(\vec{p})=\langle 1| V(\vec{p})\rangle ^*\langle 1| U(-\vec{p}\rangle ^*+\langle 2| V(\vec{p}\rangle ^*\langle 2| U(-\vec{p}\rangle ^*\nonumber\\&&
\end{eqnarray}
Where $\langle 1|$ and  $\langle 2|$ are the two orbitals.
The Hamiltonian in the chiral form  with the positive chemical potential $\mu>0$ is given by:
\begin{eqnarray}
&&H=\int\,d^3 p\Big[C^{\dagger}_{L}(\vec{p})(|\vec{p}|-\mu)C_{L}(\vec{p})
+C^{\dagger}_{R}(\vec{p})(|\vec{p}|-\mu)C_{R}(\vec{p})\nonumber\\&&+\Big(C^{\dagger}_{L}(\vec{p})C^{\dagger}_{R}(-\vec{p})F(\vec{p})-C^{\dagger}_{R}(\vec{p})C^{\dagger}_{L}(-\vec{p})F(-\vec{p})\Big)i\Delta(d)\sin(\vec{p}\cdot \vec{d})+h.c.+\frac{|\Delta(d)|^2}{2|V(d)|}\Big]\nonumber\\&&
\end{eqnarray}
We construct  the $BDG$ Hamiltonian in terms of the spinors $\Big[C_{L}(\vec{p}),C_{R}(\vec{p}),C_{R}^{\dagger}(-\vec{p}),-C_{L}^{\dagger}(-\vec{p})\Big]^{T}$.
The pairing order parameter is  represented by  $\Delta(\vec{p})=i\Delta(d)\sin(\vec{p}\cdot \vec{d})F(\vec{p})$ which obeys 
$\Delta(\vec{p})=-\Delta(-\vec{p})$ ,
$\Delta(\vec{p})=|\Delta(\vec{p})|e^{i\varphi(\vec{p})}$ with $\varphi(-\vec{p})=\varphi(\vec{p})+\pi$.

\noindent
From Eq.$(27)$   in the previous section we have:
 $C^{\dagger}_{L}(\vec{p})\Big(\partial_{i}-i a^{L}_{i}(\vec{p})\Big)C_{L}(\vec{p})$;
$C^{\dagger}_{R}(\vec{p})\Big(\partial_{i}-i a^{R}_{i}(\vec{p})\Big)C_{R}(\vec{p})$.

\noindent
We perform a gauge transformation where $P$ represents the path order operator
$C_{L}(\vec{p})=P\mathbf{\Big[e^{i\int_{p_{0}}^{\vec{p}} dp'^{i}a^{L}_{i}(\vec{p'})}\Big]}\hat{ C}_{L}(\vec{p})$ ,
$C_{R}(\vec{p})=P\mathbf{\Big[e^{i\int_{p_{0}}^{\vec{p}} dp'^{i}a^{R}_{i}(\vec{p'})}\Big]}\hat{ C}_{R}(\vec{p})$ The phase of the order parameter transform   as $\varphi(\vec{p})\rightarrow \varphi(\vec{p})-\Lambda(\vec{p})$.
\noindent
As a result  the gauge field transforms  as  $(\vec{a}^{L}(\vec{p})- \vec{a}^{R}(-\vec{p})\rightarrow\vec{a}^{L}(\vec{p})- \vec{a}^{R}(-\vec{p})-\vec{\partial}_{p} \Lambda(\vec{p})$ . Therefore the  pairing phase obeys the equation : 

\noindent
  $\vec{\partial}_{p} \varphi(\vec{p})- \Big(\vec{a}^{L}(\vec{p})- \vec{a}^{R}(-\vec{p})\Big)$ where  $\Big(\vec{a}^{L}(\vec{p})- \vec{a}^{R}(-\vec{p})\Big)\equiv \vec{A}^{(pairing)}(\vec{p};-\vec{p})$.
$\vec{A}^{(pairing)}(\vec{p};-\vec{p})$ obeys:
$\oint curl\vec{A}^{(pairing)}(\vec{p};-\vec{p})\cdot d\vec{S}=4\pi q$ with the monopole   charge $q=1$.

\noindent
The order parameter is given by: $\Delta(\vec{p})=|\Delta(\vec{p})|e^{i\varphi(\vec{p})}$ , with the phase $\varphi(\vec{p})$  to be  determined from the  solution  $\vec{\partial}\varphi(\vec{p})=\vec{A}^{(pairing)}(\vec{p};-\vec{p})$. Due to the multivalued of the pairing potential the phase can not be determined!

\noindent
 The monopole charge allows  for an expansion in terms of  a set of complete eigenfunction given by   $Y_{q,l,m}(\theta(\vec{p}),\phi(\vec{p})) $ \cite{Yang}.
For a monopole  it is not possible to choose a single vector potential in the entire space, one choses a vector potential in one sector. The complementary section is obtained by the gauge transformation  $e^{iq\phi}$. As a result also the order parameter must be defined as a section.
The \textbf{Monopole Harmonics Theorem} allows to expand  any function which obeys the section condition  \cite{Yang}:
 $|\Delta(\vec{p})|e^{i\varphi(\vec{p})}=\sum_{l,m}a_{q=1,l,m}(|\vec{p}|)Y_{q=1,l,m}(\theta(\vec{p}),\phi(\vec{p}))$
where  the complete set of eigenfunctions  $Y_{q,l,m}(\theta(\vec{p}),\phi(\vec{p})) $ are given by,
\begin{equation}
Y_{q,l,m}(\theta(\vec{p}),\phi(\vec{p}))=M_{q,l,m}\Big(1-cos(\theta)\Big)^{(-q-m)}\Big(1+cos(\theta)\Big)^{(q-m)}P_{l+m}^{(-q-m),(q+m)}(\cos(\theta))e^{i(m+q)\phi}
\label{equation}
\end{equation}
From  the   formula  for $Y_{q,l,m}(\theta(\vec{p}),\phi(\vec{p}))$ we find:
\begin{eqnarray}
&&Y_{q=1, l=0,m=0}[\theta(\vec{p}),\phi(\vec{p})]\propto \sin(\theta)e^{i\phi}, 
Y_{q=1, l=0,m=1}[\theta(\vec{p}),\phi(\vec{p})]\propto \cos^2(\theta)e^{i\phi},\nonumber\\&&
Y_{q=1, l=0,m=-1}[\theta(\vec{p}),\phi(\vec{p})]\propto \sin^2(\theta)e^{i2\phi}\nonumber\\&&
\end{eqnarray}
This shows that the order parameter $\Delta(\vec{p})=|\Delta(\vec{p})|e^{i\varphi(\vec{p})}=\sum_{l,m}a_{q=1,l,m}(|\vec{p}|)Y_{q=1, l,m}[\theta(\vec{p}),\phi(\vec{p})] $ is  different from the $p-wave$ order parameter 

\textbf{c) The effect of a scattering potential in a Wey material with broken time reversal symmetry}

\vspace{0.1 in}

Here we observe the advantage of our procedure given in Eqs.$(9,12)$ for a scattering potential $V(\vec{x})$.
\begin{eqnarray}
&&H^{ext.}=\sum_{a} \int\, d^3 q  V(\vec{q}))(-iq^{a})  \int\, d ^{3}p \Psi^{\dagger} _{R}(\vec{p})\Big[i\partial_{a}+A_{a}(\vec{p};k_{r})\Big]\Psi_{R}(\vec{p})\nonumber\\&&+\sum_{a} \int\, d^3 q  V(\vec{q}))(-iq^{a})  \int\, d ^{3}p \Psi^{\dagger} _{L}(\vec{p})\Big[i\partial_{a}+A_{a}(\vec{p};-k_{r})\Big]\Psi_{L}(\vec{p})\nonumber\\&&
\end{eqnarray}
Where $ \vec{A}(\vec{p};\vec{k}_{r})$ is the monopole vector potential and  $ \vec{A}(\vec{p};-\vec{k}_{r})$ is the antimonopole vector potential.

\noindent
For a monopole  it is not possible to choose a single vector potential in the entire space, one choses a vector potential in one sector. The complementary section is obtained by the gauge transformation  $e^{iq\phi}$  \cite{Yang} where the charge of the monopole antimonopole  is $q=\pm\frac{1}{2}$. As a result also the order parameter must be defined as a section. A theorem named  \textbf{Monopole Harmonics Theorem} says that  any function which obeys the section  condition can be expanded in the  complete set of Monopole harmonics. 
The complete set is given by :

\begin{eqnarray}
&&Y_{q=\pm\frac{1}{2},l,m}(\theta(\vec{k}),\phi(\vec{k}))
=M_{q,l,m}\Big(1-cos(\theta)\Big)^{(-q-m)}\Big(1+cos(\theta)\Big)^{(q-m)}P_{l+m}^{(-q-m),(q+m)}(\cos(\theta))e^{i(m+q)\phi}\nonumber\\&&
\end{eqnarray}


\noindent
$P_{l+m}^{(-q-m),(q+m)}(\cos(\theta))$ are the Jacoby  polynomials \cite{ Yang}

As a result we have the expansion for the spinors $\Psi_{R}(\vec{k})$ and $\Psi_{L}(\vec{k})$.

\begin{eqnarray}
&&\Psi_{R}(\vec{k})=\sum_{l,m}a_{q=\frac{1}{2},l,m}(|\vec{k}|)Y_{q=\frac{1}{2},l,m}(\theta(\vec{k}),\phi(\vec{k}));\hspace{0.1 in}\Psi_{L}(\vec{k})=\sum_{l,m}a_{q=-\frac{1}{2},l,m}(|\vec{k}|)Y_{q=-\frac{1}{2},l,m}(\theta(\vec{k}),\phi(\vec{k}))\nonumber\\&&
\end{eqnarray}
This representation can be used for computing the scattering effect in the presence of disorder. The advantage being that using this expansion we do not need to work with the singular vector potentials e $ \vec{A}(\vec{p};\pm\vec{k}_{r})$

\vspace{0.2 in}

 \textbf{VIII-Surface states  for   an arbitrary crystal-face boundary}

\vspace{0.2 in}

 Photoemmision, photoconductivity, optical conductivity   and scanning tunneling microscopy      are sensitive  to the nature    of the surface states. The topology of the surface states is affected by the physical boundary. For an arbitrary surface we  will  use a  curved coordinate   basis.  When the coordinate system rotates from point to point  we can use a  \textbf{non-coordinate basis}, $\partial_{a}$, $a=1,2,3$ introduced by Cartan \cite{Nakahara,Nieh} which is related to   the Cartesian system  by the vector  $\partial_{\mu}$ , $\mu=x,y,z$.
For example a point on a surface  is given by the cartesian coordinates  $\vec{r}=[x(u_{1},u_{2}),y(u_{1},u_{2}),z(u_{1},u_{2})]$  where       $ (u_{a=1},u_{b=2})$  are the coordinate on the surface  defined by   the  normal    $ E_{N}$ to the surface defined   by  $(u_{1},u_{2})$.  The  two sets of coordinates  are related  by   the matrix  $E^{\mu}_{a}=  \frac{\partial r^{\mu}}{\partial u_{a}}$, $\mu=x,y,z$ and $a=1,2$.  The normal to the surface  is given by  $ E_{N}=\frac{\frac{\partial \vec{r}}{\partial u_{1}} \times \frac{\partial \vec{r}}{\partial u_{2}}}{||\frac{\partial \vec{r}}{\partial u_{1}} \times \frac{\partial \vec{r}}{\partial u_{2}}||}$.
This set of transformations  allows to replace $\sigma^{a}\partial_{a}$  by the covariant derivative $\nabla_{a}= \partial_{a}+ \frac{1}{8}\Gamma^{b,c}_{a}[\sigma_{b},\sigma_{c}]$  (which  depends on the connection defined below) ; 
$ \sigma^{a}\nabla_{a}=\sigma^a E^{\mu}_{a}\Big[\partial_{\mu}+\frac{1}{8}\Gamma^{a,b}_{c}e^{c}_{\mu}[\sigma_{a},\sigma_{b}]\Big]$, $E^{\mu}_{a}$ and $ e^{c}_{\mu}$  are the transformation  and the inverse transformation matrix and  $ \Gamma^{a,b}_{c}$ is  the \textbf{connection one form} matrix (see \cite{Nakahara}  page 285).

\noindent
 The  Weyl Hamiltonian in   cylindrical coordinates  has been consideredin the literature   \cite{Ashvin,hanaguri, Franzaxion,Peng,Pnueli,DunghaiLee}. Here we will modify the  method in order to deal with the  arbitrary crystal face boundary.

\noindent  
 \textbf{a) The Weyl Hamiltonian in Cartesian coordinates  is} :$ H^{2d}=(-i)[\sigma_{x}\partial_{y}- \sigma_{y}\partial_{x}]$.

\noindent  
We put the Hamiltonian on a cylinder and take the axes $x=x^{1}$;  $y=x^{2}= r \sin[\phi]$  and  $z=x^{3}= r \cos[\phi]$;
$\vec{r}=\Big[x, r \sin[\phi],r \cos[\phi]\Big]$.
 We will  study this problem using  the \textbf{non-coordinate basis}   \cite{Nakahara}: $u_{ a=1}=x$ ,$ u_{ a=2}=\phi$ and $ r$ is the coordinate in the normal direction $E_{N}$; the \textbf{derivatives} are 
$\hat{\partial}_{a=1}=\partial_{x}$;
$\hat{\partial}_{a=2}=\frac{1}{r}\partial_{\phi}$;$\hat{\partial}_{N=3}=\partial_{r}$ and the \textbf{differentials one form}  are given by:
$\theta^{1}=dx$;  $\theta^{2}=r d\phi$;  $\theta^{3}=dr$.
The coordinate basis is not fixed,  therefore  \textbf{ connections}   $\Gamma^{a}_{a,b}$  will be generated.
The  connections $\Gamma^{a}_{a,b}$  are  obtained   from  the Cartan's  equations. We have   for the \textbf{Torsion} $T^a$ the equation  : $d \theta^{a}+\omega^{a}_{b}\wedge \theta^{b}=T^{a}$, $a=1,2,3$. The connection $ \omega^{a}_{b} $ is expanded in terms of the differential $\Gamma^{a}_{b,c}\theta^{c}$ one form using   the matrix $e^{c}_{\mu}$:

  $  e^{c}_{\mu}E^{\mu}_{a}=\delta^{a}_{c}$ ;$E^{\mu}_{a}\equiv\partial_{a} 
\vec{r}=
\Big[\partial_{a}x,\partial_{a}r \sin[\phi],\partial_{a}r \cos[\phi]\Big] $ ; 

$E^{\mu}_{1}=\Big[1,0,0\Big]$;
$E^{\mu}_{2}=\Big[0,r \cos[\phi],-r\sin[\phi]\Big]$;$E^{\mu}_{2}=\Big[0,\sin[\phi], \cos[\phi],\Big]$

\noindent
  The transformation  will determine the form of the Hamiltonian in a rotated basis :
$\omega^{a}_{b}$  and $\omega^{a,b}_{\mu}=\Gamma^{a,b}_{c}e^{c}_{\mu};
\hspace{0.1 in}\omega^{a,b}_{\mu}=-\omega^{b,a}_{\mu}$

\noindent
 From the torsion condition    $\textbf{Torsion}=T^{a} =0$   we find:
$d \theta^{a}+\omega^{a}_{b}\wedge \theta^{b}=T^{a}=0$,   ( $\wedge $ is the wedge product \cite{Nakahara}) $a=1,2,3$.  We obtain the      equation:
$d \theta^{a}=-\Gamma^{a}_{b,c}\theta^{c}\wedge\theta^{b}$
;$ \Gamma^{2}_{2,3}=-\Gamma^{2}_{3,2}=\frac{1}{r}$.

\noindent
 As a result  \textbf{the Weyl Hamiltonian in the cylindrical basis} is given by:
\begin{equation}
H^{cyl.}=    \sum_{a=1,2,3}(-i)\sigma^{a} \nabla_{a}\equiv  \sum_{a=1,2,3}(-i)\sigma^{a}\Big[\hat{\partial}_{a}+\frac{1}{8}\Gamma^{a}_{a,b}[\sigma_{a},\sigma_{b}]\Big]=
(-i)[\sigma^{1}\partial_{x}+\sigma^{2}\frac{1}{L}\partial_{\phi} - \frac{1}{2 r}\sigma^{3}]
\label{Cartan}
\end{equation}

\noindent
 The eigenvalue equation $H^{cyl.}\psi(x,\phi)=E \psi(x,\phi)$ has  real solutions for  boundary conditions $\psi(x,\phi+2\pi)=-\psi(x,\phi)$.  We find: $E=\pm\sqrt{k^2_{x}+\frac{l(l+1)}{r^2}}$;
$\psi^{\pm}(x,\phi)=e^{i k_{x}x +(l+\frac{1}{2})\phi}\Big[1,e^{-i[\pm\chi(k_{x},l)-\alpha(k_{x},l)]}\Big]^{T}$,  $k_{x}=\frac{2\pi}{L}n$ $,n=0,\pm1,\pm2,..$;     $ l= 0,\pm1,\pm2,..$ ;$ tan[\chi(k_{x},l)]=\frac{l+\frac{1}{2}}{k_{x}r}$;  $tan[\alpha(k_{x},l)]=\frac{1}{2r\sqrt{k^2_{x}+\frac{l(l+1)}{r^2}}}$.

\noindent
 \textbf{b)   The Hamiltonian  for  the three dimensional  $T.I.$  with a   cylindrical  boundary}

 We will consider   a three dimensional  $T.I.$    with the mass dependent gap  is in the $z$ direction. Such a $T.I.$  has a surface boundary   perpendicular to the  axes $z$ with crystal-face $[x,y]$ plane  localized at $z=L$.
To simplify the discussion,  we consider a situation where the crystal-face   is  cylindrical with the  cylindrical  axis  in the $ x$ direction and    length $L_{x}>L$.  As a result, any  point on the surface of the cylinder is given  by the  set of coordinates   $\vec{r}=[x,L\sin(\phi),L\cos(\phi)]$.

The four-band model for the three dimensional $T.I.$, $Bi_{2}Se_{3}$  \cite{Kane}   of size $L_{x}\times L\times L$  
is given by, $ H=H_{\bot}+H_{\|}$ where \textbf{ $\vec{\tau}$  are the Pauli matrix for the orbitals} and \textbf{ $\vec{\sigma}$   describes  the spin}:
\begin{equation}
H_{\bot}=\tau_{3}(-m_{0}(z)-m_{2}\partial^2_{z}) +i\tau_{2}\partial_{z}, \hspace{0.1 in}
H_{\|}=(-i)\tau_{1}(\sigma_{2}\partial_{y}- \sigma_{1}\partial_{x})-m_{\|}\tau_{3}(\partial^2_{x}+\partial^2_{y})
\label{bot}
\end{equation}

\noindent 
The mass is a function  of $z$,  $ m_{0}(z)=|m_{o}|\Theta[-z+L]+(-M)\Theta[z-L]$,  where $\Theta[z]$ is the step function.  For $ z>L$, we consider $ M\rightarrow \infty$, therefore  we obtain a  
 zero mode on the surface.
 For  a crystal-face which is cylindrical,  any point on the cylinder  makes an angle of   $\phi$  with  the $z$ axis.

\noindent 
We consider a situation where the surface perpendicular to  axis $z$  (the direction of the mass gap is   given by  $H_{\bot}$) is is a surface of a cylinder. Any point on the surface can be viewed as rotated   by an angle $\phi$ (the  new axis   $z'$ makes an angle   $\phi$  with the original axis $z$ for which the mass gap has been introduced ) The axes $z'$  becomes the radial direction on a cylinder.  
 The  transformed  Hamiltonian $ H^{'}_{\|}$  is expressed in terms of the covariant derivative  $\nabla_{a}$. $ H^{'}_{\bot}$  and  $H^{'}_{\|}$  are given by  :
\begin{eqnarray}
&&H^{'}_{\bot}=\tau_{3}[-m_{0}(\frac{r}{\cos(\phi)})-m_{2} (\frac{r}{\cos(\phi)}) \cos^2(\phi)\partial^2_{r} ]+[i\tau_{2}\cos(\phi)-i\tau_{1}\sigma_{x } \sin(\phi)\partial_{r}]\nonumber\\&&
H^{'}_{\|}=(-i)[\tau_{1}\sigma_{x}\cos(\phi)(-\nabla_{2})+\tau_{1}\sigma_{y} (\nabla_{1})]\equiv i\tau_{1}[\sigma_{x}\frac{\cos(\phi)}{r}\partial_{\phi}+ \sigma_{y}\partial_{x}+\frac{\sigma_{z}}{2r}]\nonumber\\&&
\end{eqnarray}
\noindent 
The Hamiltonian $ H^{'}_{\bot}$  has  zero mode solutions  $ \hat{\Omega}(\vec {r})$, 
$ H^{'}_{\bot}\hat{\Omega}(\vec{r})=0$. The   zero mode spinor     $\hat{\Omega}(\vec{r})=\eta_{\tau,\sigma}\Omega(\vec{r})$  is given by a  product of  the  scalar function  $\Omega(\vec{r})$   and  the spinor $\eta_{\tau,\sigma} $ . The  scalar function is localized at $r=L$  and vanishes for  $ r  \rightarrow\infty$. Using the eigenvalues of $\sigma^{x}$ and $ \tau_{1}$   we find:

\noindent
$\eta_{\tau=1,\sigma; cos(\phi)>0}=\frac{1}{\sqrt{2}}[1,e^{i \sigma\phi}]\otimes\frac{1}{\sqrt{2}}[1,\sigma]$; 
$ \eta_{\tau=-1,\sigma; cos(\phi)<0}=\frac{1}{\sqrt{2}}[1, -e^{i \sigma\phi}]\otimes\frac{1}{\sqrt{2}}[1,\sigma]$ ;  $\sigma=\pm 1$

\noindent
  We define the  vectors      $|1>=\frac{1}{\sqrt{2}}[\tau=1,\sigma=1] $, $|-1>=\frac{1}{\sqrt{2}}[\tau=1,\sigma=-1]$  which allow to introduce  the rotated   Pauli matrices :

 $S_{3}= \frac{1}{2}[|1\rangle\langle 1|- |-1\rangle\langle-1|]$,  $S_{1}=\frac{1}{2}[ |1 \rangle\langle-1|+  |-1\rangle\langle+1|]$  and      $S_{2}= \frac{1}{2}[-i|1\rangle\langle-1|+i  |-1\rangle\langle+1|]$ 

\noindent
 We will use the  eigenstates $\eta_{\tau=1,\sigma; cos(\phi)>0}$ to  compute   the projections   $<\tau_{1}\sigma_{x}>|_{\tau=1} $ $<\tau_{1}\sigma_{y}>|_{\tau=1}$ for the Hamiltonian $H^{'}_{\|}=(-i)[\tau_{1}\sigma_{x}\cos(\phi)(-\nabla_{2})+\tau_{1}\sigma_{y} (\nabla_{1})]$ (see Eq.$(48)$)

We find:

  $<\tau_{1}\sigma_{x}>|_{\tau=1}=S_{3}\cos(\phi)$ ;$<\tau_{1}\sigma_{y}>|_{\tau=1}=(-S_{2}\cos(\phi)+S_{1}\sin(\phi))$ 

As a result  the Hamiltonian $H^{'}_{\|}$ depends on the  angle $\phi$ :  
\begin{equation}
 H^{\cos(\phi)}_{\|}=(-i)\Big[-\frac{cos^2(\phi)}{r}S_{3}\partial_{\phi}+ \cos(\phi)
(S_{2}-S_{1}\sin(\phi))\partial_{x}-\frac{1}{r}( \cos(\phi)S_{1}+S_{2}\sin(\phi))]
=-H^{\cos(\phi+\pi)}_{\|}
\label{final}
\end{equation}
\noindent
  We observed that the rotation of the crystal-face for a cylinder is different from the surface state of  a cylinder. In particular We observe  a    change  of sign of the Hamiltonian    for a  large angles, $ H^{\cos(\phi)}_{\|}=-H^{\cos(\phi+\pi)}_{\|}$. 

\noindent
To conclude,  the eigenvectors  
  determines    the  surface properties,  such as  spin texture,  and surface currents . We believe that  this results can be observed   by   photoemmision or  hotoconductivity.

\vspace{0.2 in}

  \textbf{IX-Topological invariant for Superconductors}

\vspace{0.2 in}

For superconductors we can use the invariance in the entire B.Z.  obtained by combining the time reversal invariance and the particle-hole symmetry. As a result one can find a unitary  matrix $\Gamma$,  $\Gamma^2=1$ which anti commutes with the Superconductor Hamiltonian.  One can show that the Hamiltonian can be brought to the form ( due to the Superconducting gap we can use a flat Hamiltonian):
\begin{eqnarray}
&&Q(\vec{k})=\left[\begin{array}{rrr}
                                       0 &q(\vec{k}) \\
q^{\dagger}(\vec{k}) & 0 \\
\end{array}\right] 
\nonumber\\&&
\end{eqnarray}
\noindent From the relation $ch_{2}=d[K_{3}(A,F)]$    we identify the winding number $\nu_{3}=\int_{  S^{4}} d[K_{3}(A,F)]$.
\begin{eqnarray}
&&\nu_{3}=\int_{  \partial S^{4}}(K_{3}[\mathcal{A}_{+},\mathcal{F}_{+}]-K_{3}[\mathcal{A}_{-},\mathcal{F}_{-}])
\nonumber\\&&=
 \frac{1}{24\pi^2}\int\,d^{3}k \epsilon_{i,j,k}Tr[(q^{-1}(\vec{k})\partial_{i}q^{\dagger}(\vec{k}))(q(\vec{k})\partial_{j}q^{-1}(\vec{k}))(q(\vec{k})\partial_{k}q^{-1}(\vec{k}))]\nonumber\\&&
\end{eqnarray}
\noindent In one dimensions we have $\nu_{1}=\frac{1}{i2\pi}\int_{B.Z.}dk Tr[q(\vec{k})\partial_{k}q^{-1}(\vec{k})]$.

\noindent For two dimensions one identifies the $Z_{2}$ index with the Pfaffian matrix   (the Pfaffian at the time reversal invariant point is equal  to the matrix elements  of the Hamiltonian components $q(\vec{k})$  \cite{Ludwig,Schneider,Ludwigg}):
\begin{equation}
I=\prod_{\vec{\Gamma}}\frac{\mathbf{Pf}\Big[q^{T}[\vec{\Gamma}]\Big]}{\sqrt{det[q[\vec{\Gamma}]]}}
\label{index}
\end{equation} 
See additional discussions in Appendix B.

\vspace{0.2 in}

 \textbf{X-Superconductivity on a T.I. surface}

\vspace{0.2 in}
. 
We consider a situation where the chemical potential  $\mu >0$  obeys $\mu>\Delta$ (the pairing field). Following \cite{FuKane} we have for the the projected Hamiltonian, 
\begin{equation}
H=\sum_{k}(v|\vec{k}|-\mu)c^{\dagger}(\vec{k}) c(\vec{k})+(\Delta c^{\dagger}(\vec{k})e^{i\theta(\vec{k})}c^{\dagger}(\vec{k}) c^{\dagger}(-\vec{k})+h.c.) 
\label{hamiltonian}
\end{equation}
where $e^{i\theta(\vec{k})} =\frac{k_{1}-ik_{2}}{k}$. For the  conduction  electron  the surface eigenvectors  are :  $|u^{+}(\vec{k}) \rangle =\frac{1}{\sqrt{2}}\Big[1,\frac{k_{1}-ik_{2}}{k}F(\frac{\vec{k}}{k_{0}})\Big]$  gives rise in  the momentum space to a  vortex, $F(\frac{\vec{k}}{k_{0}})$  given by  $F(\frac{\vec{k}}{k_{0}})=1-e^{-\frac{|\vec{k}|}{k_{0}}}$.
 The Hamiltonian in the  long wave limit is obtained by replacing  the kinetic energy $ \hbar v (|\vec{k}|-\vec{k}_{F})\approx  \frac{\hbar v}{2 k_{F}}[k^2-k^2_{F}]$. The term   $\frac{(k_{1}+ik_{2})}{k} F(\frac{\vec{k}}{k_{0}})$ is replaced  by $\frac{(k_{1}+ik_{2})}{k_{0}}$  which is   a  spatial derivatives in real space.
The  gauge transformation in the presence of a \textbf{superconducting  vortex}  is:
$\Delta (\vec{r})\rightarrow \Delta_{0} e^{i\theta(\vec{r})}$ ; $\Delta^{*} (\vec{r})\rightarrow \Delta_{0} e^{-i\theta(\vec{r})}$;
$C(\vec{r},t) \rightarrow  e^{\frac{-i}{2}\theta(\vec{r})}
 \hat{C}(\vec{r},t)$;
$C^{\dagger}(\vec{r},t) \rightarrow \hat{C}^{\dagger}(\vec{r},t) e^{\frac{-i}{2}\theta(\vec{r},t)}$.
In the polar representation  the field  $\hat{C}(\vec{r})=\hat{C}(r,\phi)$  and  $\hat{C}^{\dagger}(\vec{r})$ transform in the presence  of the vortex phase  $\theta(\vec{r})$,  $\hat{C}(r,\phi +2\pi)= -\hat{C}(r,\phi)$ ,$\hat{C}^{\dagger}(r,\phi +2\pi)= -\hat{C}^{\dagger}(r,\phi)$. Consequently for  a $2\pi$ 
vortex the fields $\hat{C}(\vec{r}) $ ,$\hat{C}^{\dagger}(\vec{r})$ are \textbf{double-valued}.

\noindent 
The chemical potential  $\mu_{eff.}$ in the presence of a superconducting  vortex   at $\vec{r}\rightarrow 0$ changes the  sign.
\begin{eqnarray}
&&H=
\int d^2r \Big[\hat{C}^{\dagger}(r,\phi)[\frac{\hbar v}{2 k_{F}}(-\partial ^2_{r}-\frac{1}{r^2}\partial^2_{\phi})-\mu_{eff.}(\vec{r})]\hat{C}(r,\phi)
-\frac {|\Delta| }{2k_{0}}\hat{C}(r,\phi)e^{i\phi}(\partial_{r}+\frac{i}{r}\partial_{\phi})\hat{C}(r,\phi) \nonumber\\&&    +    \frac {|\Delta|                  }{2k_{0}} \hat{C}^{\dagger}(r,\phi)e^{-i\phi}(\partial_{r}-\frac{i}{r}\partial_{\phi})\hat{C}(r,\phi) \Big] +\frac{1}{g}\int dt\int d^2r |\Delta(\vec{r},t)
\nonumber\\&&
\end{eqnarray}
\noindent Folowing \cite{Ivanov,Taylor} we observe that this Hamiltonian has a zero mode Majorana solution   $\Big[U(r,\phi),V(r,\phi)\Big]$ : 
$\Big[U(r,\phi),V(r,\phi)\Big]^{T}\equiv\Big[\frac{1}{\sqrt{i}}e^{\frac{i }{2}\phi},\frac{1}{\sqrt{-i}}e^{\frac{-i }{2}\phi}\Big]^{T}\frac{f(r)}{\sqrt{r}}$.
The function $\frac{f(r)}{\sqrt{r}}$ obeys the normalization condition, $\int\,d^2 r [\frac{f(r)}{\sqrt{r}}]^{2}<\infty$.

\noindent
 Next  we  consider \textbf{few vortices}, we   limit ourself to $N=1$ (odd vortices) localized at  the space  time  coordinates $ [\vec{R}_{1}(t)],....[\vec{R}_{2n}(t)]$ . At these points  the amplitude   pairing field  is  $|\Delta(\vec{r},t)|$. We consider  a situation where  the  $2n$ vortices  are  far from each other.  Each vortex corresponds to one Majorana Fermion \cite{Ivanov}$\gamma_{i},i=1,..2n$.   
$\Delta(\vec{r},t,)=|\Delta(\vec{r},t,)|e^{i\theta(\vec{r},t)}\approx \Delta_{0} 
e^{i\theta_{0}(\vec{r},t)} e^{i\theta_{v}(\vec{r},t)}$
$\theta_{v}(\vec{r},t)\equiv \theta_{v}(\vec{r},t;\vec{R}_{1}(t),....\vec{R}_{2n}(t)$.
 The phase  $\theta_{v}(\vec{r},t)$  satisfies :
$J_{\mu}(\vec{r},t)=\frac{1}{2\pi}\epsilon_{\mu,\nu,\lambda}\partial_{\nu}\partial_{\mu}\theta_{v}(\vec{r},t)$,
$J_{0}(\vec{r},t)=\frac{1}{2\pi}[\partial_{1}\partial_{2}-\partial_{2}\partial_{1}]\theta_{v}(\vec{r},t)=
\sum_{\vec{R}_{i}}z_{i}\delta[\vec{r}-\vec{R}_{i}(t)]$,
$\vec{J}(\vec{r},t)=\sum_{\vec{R}_{i}}z_{i}[\partial_{t}\vec{R}{i}(t)] \delta[\vec{r}-\vec{R}_{i}(t)]$,
$z_{i}=\pm 1$.
\noindent  In order to evaluate the effect of the Majorana vortices we introduce the  Majorana density,
\begin{equation}
n(\vec{r},t)=n^{odd}(\vec{r},t)+ n^{even}(\vec{r},t)=\sum_{a=1}^{n}\Big[\delta(\vec{r}-\vec{R}_{2a-1}(t))+\delta(\vec{r}-\vec{R}_{2a}(t)]\Big]
\label{density}
\end{equation}
 The  Majorana current  density is given by,
\begin{equation}
 \vec{n}(\vec{r},t)=\vec{n}^{odd}(\vec{r},t)+ \vec{n}^{even}(\vec{r},t)= \sum_{a=1}^{n}\Big[\partial_{t}\vec{R}_{2a-1}(t)\delta(\vec{r}-\vec{R}_{2a-1}(t))+\partial_{t}\vec{R}_{2a}(t)\delta(\vec{r}-\vec{R}_{2a}(t))\Big]
\label{current}
\end{equation}
The Berry phase involves the time derivative of the Majorana operators  $\hat{\gamma}_{2a-1}(t)\equiv\hat{\gamma} (\vec{R}_{2a-1}(t),t)$ and $\hat{\gamma}_{2a}(t)\equiv\hat{\gamma} (\vec{R}_{2a}(t),t)$.
We compute   the time derivative $\frac{d}{dt}\hat{\gamma}(\vec{R}_{2a-1}(t),t)$ and  $\frac{d}{dt}\hat{\gamma}(\vec{R}_{2a}(t),t)$ which  are defined with respect the moving coordinates  $\vec{R}_{2a-1}(t)$, $\vec{R}_{2a}(t)$ (the position of the  vortices). We  construct  the  dual action  introduced in ref. \cite{Arovas}:
\begin{eqnarray}
&&S=\int\,dt\int\,d^2 r \hat{\gamma}(\vec{r},t)\Big[i n(\vec{r},t)\partial_{t}+i \vec{n}(\vec{r},t)\cdot\vec{\partial_{r}}\Big]\hat{\gamma}(\vec{r},t)
+\Big[\frac{1}{2\pi} \epsilon^{\mu,\nu,\lambda}\partial_{\nu}b_{\lambda}(\vec{r},t)[
-e\mathbf{A}_{\lambda}(\vec{r},t)+\frac{1}{2}a^{v}_{\lambda}(\vec{r},t)] \nonumber\\&&+\frac{1}{2\pi}\epsilon^{\mu,\nu,\lambda}C_{\mu}(\vec{r},t)\partial_{\nu}a^{v}_{\lambda}(\vec{r},t)-C_{\mu}(\vec{r},t) J_{\mu}(\vec{r},t)\Big]\nonumber\\&&
\end{eqnarray}

\noindent
 In order to  evaluate the effect of the motion of the Majorana vortices we replace  the $2n$ Majorana  fermions by  $n$  spinless fermions.
\noindent The  field $C_{\mu}(\vec{r},t)$  acts as a constraint  which imposes the relation $ J_{\mu}(\vec{r},t)=\frac{1}{2\pi}\epsilon^{\mu,\nu,\lambda}\partial_{\nu}a^{v}_{\lambda}(\vec{r},t)=\frac{1}{2\pi}\epsilon_{\mu,\nu,\lambda}\partial_{\nu}\partial_{\mu}\theta_{v}(\vec{r},t)$,
  The vortices are described by $\frac{1}{2}a^{v}_{\lambda}(\vec{r},t)$. In a regular superconductor the vortices are screened and the ground state  response of the external magnetic field gives rise to the  Meisner effect. 
 
\vspace{0.2 in}

 \textbf{XI-Transport trough Majorana Fermions-Resonant  Andreev crossed reflection.}

\vspace{0.2 in}

\textbf{a)-The Majorana modes}

\vspace{0.1 in}

A direct evidence for  a  topological superconductor can be   observed   from transport. Attaching two metallic leads to  a superconductor  which support  Majorana fermions  one can observe the   resonant Andreev  reflection  \cite{Ng} and the Crossed  Majorana fermion   Andreev  transmission .   Superconductors have an Andreev reflection  $\propto  (\frac{\Gamma}{\Delta})^2$ which is smaller in comparison to the resonant  Majorana Andreev reflection which is of the order  of 1.
Most of the work done was for  reflection ( an incoming  electron is reflected as a hole and a charge of $2e$ is propagating in the Superconductor)  was    investigated with the help of the $\mathbf{S}$ matrix \cite{Beenaker,Ng,Fisher,Buttiker,Kane,Flensberg}. One of the difficulties for investigating the \textbf{resonant  Andreev crossed reflection} (an incoming electron  at lead $1$ is reflected  as an outgoing hole in lead $2$) is due to the fact that  the Majorana modes have  a finite energy. The  $\mathbf{S}$ matrix
is  obtained  by imposing the continuity equation an unitarity.
 The Majorana modes makes  it difficult to  compute the $\mathbf{S}$ matrix  and to represent it  as a Dyson series  $\mathbf{S}=T\Big[e^{- i \int_{-\infty}^{\infty}\,dt' H_{t}(t')}\Big]$  where  $H_{t}(t)=\int\,d^{d}r h(\vec{r},t)$ is the perturbing Hamiltonian.
The knowledge  of the  $\mathbf{S}$  matrix   given in terms of the Dyson equation has the advantage of allowing the use of the "mashinery" of the   Renormalization Group \cite{Weinberg,Ramond,davidimpurity}. The Majorana fermions have no coherent states,therefore the construction of the path integral looks impossible.
 This problem can be solved  by  halving the degrees of freedom and introducing  spinless fermions which .  
 For an \textbf{even} number of Majorana fermions  $ \hat{\gamma}_{2a-1}$ and  $\hat{\gamma}_{2a}$ $a=1,2,3,4...$  with the eigenfunctions $ F_{2a-1}(\vec{r})$, $F_{2a}(\vec{r})$ we have the representation for the zero mode fermions:
$\hat{C}_{0}(\vec{r})=\hat{C}^{\dagger}_{0}(\vec{r})=\sqrt{2}\sum_{a=1}^{m}\Big[\hat{\gamma}_{2a-1}F_{2a-1}(\vec{r})+\hat{\gamma}_{2a}F_{2a}(\vec{r})\Big]$,
 where  $F_{2a}(\vec{r})=
\Big[\frac{1}{\sqrt{i}}e^{\frac{i }{2}\phi_{2a}},\frac{1}{\sqrt{-i}}e^{\frac{-i }{2}\phi_{2a}}\Big]^{T}\frac{f_{2a}(r)}{\sqrt{r}}$  and $F_{2a-1}(\vec{r})=
\Big[\frac{1}{\sqrt{i}}e^{\frac{i }{2}\phi_{2a-1}},\frac{1}{\sqrt{-i}}e^{\frac{-i }{2}\phi_{2a-1}}\Big]^{T}\frac{f_{2a-1}(r)}{\sqrt{r}}$ are the  two component spinors  localized at the positions $\vec{r}=\vec{R}_{2a-1}$ and $\vec{r}=\vec{R}_{2a}$.
We introduce the spinless  fermion operators $ \zeta^{\dagger}_{a}$,  $\zeta_{a}$ , $a=1,2,3...n$  and construct the Fermionic path integral. The transformation between the two representation is given by: 
  $\hat{\gamma}_{2a-1}=\frac{1}{\sqrt{2}}\Big[\zeta^{\dagger}_{a}+\zeta_{a}\Big]$ , 
$\hat{\gamma}_{2a}=\frac{1}{i\sqrt{2}}\Big[\zeta^{\dagger}_{a}-\zeta_{a}\Big]$, $a=1,2,3...n$
\noindent The low energy   Hamiltonian takes  the form :
$S^{(Majorana)}= \sum_{a=1}^{a=n}\int\,dt  \Big[\zeta^{\dagger}_{a}(i\partial_{t})\zeta_{a}-\epsilon_{a}\zeta^{\dagger}_{a}\zeta_{a}\Big]$, $\epsilon_{a}$ is the overlapping energy for the Majorana Fermions $i\epsilon_{a}\hat{\gamma}_{2a-1}\hat{\gamma}_{2a}$.
\noindent For an \textbf{odd} number of Majorana Fermions ( we will have for the  $2 n+1$  Majorana  an unpaired Fermionic)  ,we  choose for  $\hat{\gamma}_{n+1}=\frac{1}{\sqrt{2}}\Big[\zeta^{\dagger}_{n+1}+\zeta_{n+1}\Big]$\textbf{ or}   $\hat{\gamma}_{n+1}=\frac{1}{i\sqrt{2}}\Big[\zeta^{\dagger}_{n+1}-\zeta_{n+1}\Big]$. 
\noindent
The coupling of the Fermions to the  two leads is given by 
\begin{equation}
H_{t}=t\sum_{\sigma=\uparrow,\downarrow}\int\,dy\Big[d^{\dagger}_{\sigma}(x=-\frac{L}{2},y)\Psi_{\sigma}((x=-\frac{L}{2},y)+d^{\dagger}_{\sigma}(x=\frac{L}{2},y)\Psi_{\sigma}((x=\frac{L}{2},y)+h.c.\Big]
\label{M}
\end{equation}
We consider two metallic leads at $ x=-\frac{L}{2}$ (first lead ) and $ x=\frac{L}{2}$ (second lead ). 
Keeping only the Majorana Fermions  we replace   $\Psi_{\sigma}(x,y)$ with  the zero mode  two component spinor    $\hat{C}_{0}(\vec{r})$. 
 
\vspace{0.1 in}

 \textbf{b)  The differential conductivity for  the Andreev crossed reflection-A pair of two Majorana vortices}

\vspace{0.1 in}

\noindent
Contrary to the single Majorana half vortex, the presence of  two half  vortices induces a current  trough the superconductor. We will  compute the \textbf{Crossed Andreev  reflection} which is a process in which an incoming electron from lead $1$, is turned into an outgoing hole in lead $2$. In this case a single electron at each lead   is tunneling into the superconductor to form a Cooper pair.  
 For simplicity  we consider two such vortices localized    at the edges of the superconductor with the \textbf{tunneling} Hamiltonian and  Majorana \textbf{overlapping}: 
\begin{eqnarray}
&&H_{t}=g\Big[d^{\dagger}_{1}(x=-\frac{L}{2},0)e^{i\frac{\phi_{1}}{2}}-d_{1}(x=-\frac{L}{2},0)e^{-i\frac{\phi_{1}}{2}}\Big]\hat{\gamma}_{1}+ \Big[d^{\dagger}_{2}(x=\frac{L}{2},0)e^{i\frac{\phi_{2}}{2}}-d_{2}(x=\frac{L}{2},0)e^{-i\frac{\phi_{2}}{2}}\Big]\hat{\gamma}_{2}\nonumber\\&&
H^{(Majorana)}=i\epsilon_{0}\hat{\gamma}_{1}\hat{\gamma}_{2}\nonumber\\&&
\end{eqnarray}
 The overlapping of the  two  Majorana Fermions is given by   $i\epsilon_{0}\hat{\gamma}_{1}\hat{\gamma}_{2}$ which are   localized at different position and   are  not-orthogonal. 
We replace the   Majorana  operators   with their fermionic representation  $\hat{\gamma}_{1}=\frac{1}{\sqrt{2}}\Big[\zeta^{\dagger}+\zeta\Big]$  and $\hat{\gamma}_{2} =\frac{1}{i\sqrt{2}}\Big[\zeta^{\dagger}-\zeta\Big]$ .
The $1d$ leads are  $d_{1}(x=-\frac{L}{2})$ and  $d_{2}(x=\frac{L}{2})$.
We integrate the Majorana fermions and obtain  the  $\mathbf{S}$  matrix for scattering between the leads:
\begin{eqnarray}
&&\mathbf{S}=(-ig^2) \int_{-\infty}^{\infty}\,dt\int_{0}^{\infty}\,d \tau\Big[V^{\dagger}(t)e^{i\epsilon_{0} \tau}V(t-\tau)\Big]\nonumber\\&&
V^{\dagger}(t) V(t-\tau)=\Big(d^{\dagger}_{1}(t)e^{-i\frac{\phi_{1}}{2}}-d_{1}(t)e^{i\frac{\phi_{1}}{2}}+ id^{\dagger}_{2}(t)e^{-i\frac{\phi_{2}}{2}}-id_{2}(t)e^{i\frac{\phi_{2}}{2}}\Big)\cdot\nonumber\\&&
\Big(-d^{\dagger}_{1}(t-\tau)e^{-i\frac{\phi_{1}}{2}}+d_{1}(t-\tau)e^{i\frac{\phi_{1}}{2}}+ id^{\dagger}_{2}(t-\tau)e^{-i\frac{\phi_{2}}{2}}-id_{2}(t-\tau)e^{i\frac{\phi_{2}}{2}}\Big)\nonumber\\&&
\end{eqnarray}
We expand the fermion operators in the leads using the  right and left movers.
$d_{1}(t)(x=-\frac{L}{2},0)=R_{1}(t)e^{-ik_{F}\frac{L}{2}}+
L_{1}(t)e^{ik_{F}\frac{L}{2}}$; 
$d_{2}(t)(x=\frac{L}{2},0)=R_{2}(t)e^{ik_{F}\frac{L}{2}}+
L_{2}(t)e^{-ik_{F}\frac{L}{2}}$.
We apply on the left lead a voltage  $V/2$ and on the  right lead a voltage $-V/2$.
As a result we obtain for each leads two Green's functions:
\begin{eqnarray}
&&\mathbf{G}_{0}^{1,R}(E,\omega)=\frac{\theta(E-\frac{eV}{2})}{\omega-(E-\frac{eV}{2})+i0}+\frac{\theta(-E+\frac{eV}{2})}{\omega-(E-\frac{eV}{2})-i0}\nonumber\\&&
\mathbf{G}_{0}^{1,L}(E,\omega)=\frac{\theta(-E+\frac{eV}{2})}{\omega+(E-\frac{eV}{2})-i0}+\frac{\theta(E-\frac{eV}{2})}{\omega+(E-\frac{eV}{2})+i0}\nonumber\\&&
\mathbf{G}_{0}^{2,R}(E,\omega)=\frac{\Theta(E+\frac{eV}{2})}{\omega-(E+\frac{eV}{2})+i0}+\frac{\Theta(-E-\frac{eV}{2})}{\omega-(E+\frac{eV}{2})-i0}\nonumber\\&&
\mathbf{G}_{0}^{2,L}(E,\omega)=\frac{\Theta(-E-\frac{eV}{2})}{\omega+(E+\frac{eV}{2})-i0}+\frac{\Theta(E+\frac{eV}{2})}{\omega+(E+\frac{eV}{2})+i0}\nonumber\\&&
\end{eqnarray}
$\Theta(x)$ is the step function (zero for $ x<0$ and one for $x\geq0$)
The current in the  leads is given by:
$J(x=-L;\frac{V}{2})=ev(N_{0}^{1,R}-N_{0}^{1,L})=J(x=L;\frac{-V}{2})=-ev(N_{0}^{2,R}-N_{0}^{2,L})$.
In order to compute the current  we will compute the Green's functions  as  function of the tunneling parameter and the  voltage applied on the two leads:
$\mathbf{G}^{1,R}(E,\omega;\frac{eV}{2})$, $\mathbf{G}^{1,L}(E,\omega;\frac{eV}{2})$ (left lead) and 
$\mathbf{G}^{2,R}(E,\omega;-\frac{eV}{2})$, $\mathbf{G}^{2,L}(E,\omega;-\frac{eV}{2})$ (right lead).
The   self energies for each lead are a function of the frequency ;$\mathbf{\Sigma^{1,R}}(\omega)$, $\mathbf{\Sigma^{1,L}}(\omega)$ and  $\mathbf{\Sigma^{2,R}}(\omega)$, $\mathbf{\Sigma^{2,L}}(\omega)$.


\noindent 
We find, to order  $g^4$,  the self energies:
\begin{eqnarray}
&&\mathbf{\Sigma^{1,R}}(\omega)=-2T(\omega,\omega_{0})\frac{g^4}{v}Ln\Big(\frac{1+\frac{\omega-\frac{eV}{2}}{\Lambda}}{1-\frac{\omega-\frac{eV}{2}}{\Lambda}}\Big)+i2T(\omega,\omega_{0})\frac{g^4}{v}\mathbf{sgn}(\omega)\nonumber\\&&
\mathbf{\Sigma^{1,L}}(\omega)=-2T(\omega,\omega_{0})\frac{g^4}{v}Ln\Big(\frac{1+\frac{\omega-\frac{eV}{2}}{\Lambda}}{1-\frac{\omega-\frac{eV}{2}}{\Lambda}}\Big)-i2T(\omega,\omega_{0})\frac{g^4}{v}\mathbf{sgn}(\omega)\nonumber\\&&
T(\omega,\omega_{0})=\frac{1}{\Gamma^2_{0}+(\omega+\omega_{0})^2}\nonumber\\&&
\end{eqnarray}
Where $\Lambda$ is the band with, $\Gamma_{0}$ is a damping factor which is and   $\hbar\omega_{0}=\epsilon_{0}$ is the Majorana energy. The \textbf{imaginary} part of the self energy obeys \textbf{ $Im. \mathbf{\Sigma^{1,L}}(\omega)=-Im. \mathbf{\Sigma^{1,R}}(\omega)$} and the \textbf{real part of the self energy} obeys \textbf{ $\Re \mathbf{\Sigma^{1,L}}(\omega)=\Re \mathbf{\Sigma^{1,R}}(\omega)\equiv\mathbf{\Sigma^{1}}(\omega)$}.
The Green's funtions are  given in terms of the  self energies:
\begin{eqnarray}
&&\mathbf{G}^{1,R}(E,\omega;\frac{eV}{2})=\Big(\omega-(E-\frac{eV}{2})-\mathbf{\Sigma^{1,R}}(\omega)\Big)^{-1}\nonumber\\&&
\mathbf{G}^{1,L}(E,\omega;\frac{eV}{2})=\Big(\omega+(E-\frac{eV}{2})-\mathbf{\Sigma^{1,L}}(\omega)\Big)^{-1}\nonumber\\&&
\end{eqnarray}
The real part of the self energy is used to compute the wave function renormalization function  $\mathbf{Z}$.
\begin{eqnarray}
&&\Big(1-\partial_{\omega}\mathbf{\Sigma^{1}}(\omega)\Big)|_{\omega=0}=\mathbf{Z}^{-1}\nonumber\\&&
\mathbf{Z}^{-1}=\Big[1+\frac{\hat{\Gamma}}{\Lambda}\Big(\frac{1}{1-\frac{eV}{2\Lambda}}\Big]\nonumber\\&&
\hat{\Gamma}=\frac{4T(\omega=0,\omega_{0})g^4}{v }= \frac{4g^4}{v (\Gamma^2_{0}+\omega_{0}^2)}\nonumber\\&&
\end{eqnarray}
$\hat{\Gamma}$ is the microscopic width  defined at the short lenght scale $l=log\Big[ \frac{\Lambda}{eV}\Big]=0$ given in Eq.$(64)$.

\noindent 
Due to the nonlinearity of the effective action Eq.$(60)$  the width $\Gamma$  scales with the voltage   $\Gamma\Big[l=log\Big(\frac{\Lambda}{eV}\Big)\Big]$ .   Using scaling equation   \cite{Ramond,Weinberg,Boyanovsky,daviddirac,davidimpurity}  for the coupling constant $g^2$ we find that the   the width satisfies the equation:

\noindent
$\frac{d{\Gamma}}{dl}=-const.{\Gamma}^{2}$ with $ l=log\Big[ \frac{\Lambda}{eV}\Big]$

\noindent
 We obtain    the \textbf{differential conductance for the Crossed Andreev reflection} $\frac{d I(V)}{dV}$  :
\begin{eqnarray}
&&\frac{d I(V)}{d V}=\frac{e^2}{h}\int_{-\Lambda \mathbf{Z}}^{\Lambda \mathbf{Z}}\,d\epsilon \int_{-\infty}^{\infty}\,\frac{d\Omega} {2\pi}\Big [\frac{\hat{\Gamma}\mathbf{Z}}{(\Omega-\epsilon)^2+(\hat{\Gamma}\mathbf{Z})^2}\Big]\frac{d}{d\Omega}\Big(f_{F.D.}(\Omega+\frac{eV}{2}\mathbf{Z})+f_{F.D.}(\Omega-\frac{eV}{2}\mathbf{Z})\Big)\nonumber\\&&
\approx  \frac{e^2}{h}\int_{-\Lambda }^{\Lambda} \,d\epsilon\frac{1}{2\pi}\Big[\frac{\hat{\Gamma}(V)}{(\frac{eV}{2}-\epsilon)^2+(\hat{\Gamma}(V))^2}+\frac{\hat{\Gamma}(V)}{(\frac{eV}{2}+\epsilon)^2+(\hat{\Gamma}(V))^2}\nonumber\\&&
=\frac{e^2}{h}\frac{1}{\pi}\Big[ArcTan[\frac{\Lambda}{\Gamma(V)}(1+\frac{eV}{2\Lambda})]+ArcTan[\frac{\Lambda}{\Gamma(V)}(1-\frac{eV}{2 \Lambda})]\nonumber\\&&
\Gamma(l(V))\equiv\Gamma(V)=\frac{\Gamma(l=0)}{(1+const.\Gamma(l=0)Log[\frac{\Lambda}{eV}])},\vspace{0.5 in} \Gamma(l=0)\equiv \hat{\Gamma}\nonumber\\&&
\end{eqnarray}
\begin{figure}
\begin{center}
\includegraphics[width=4.0 in ]{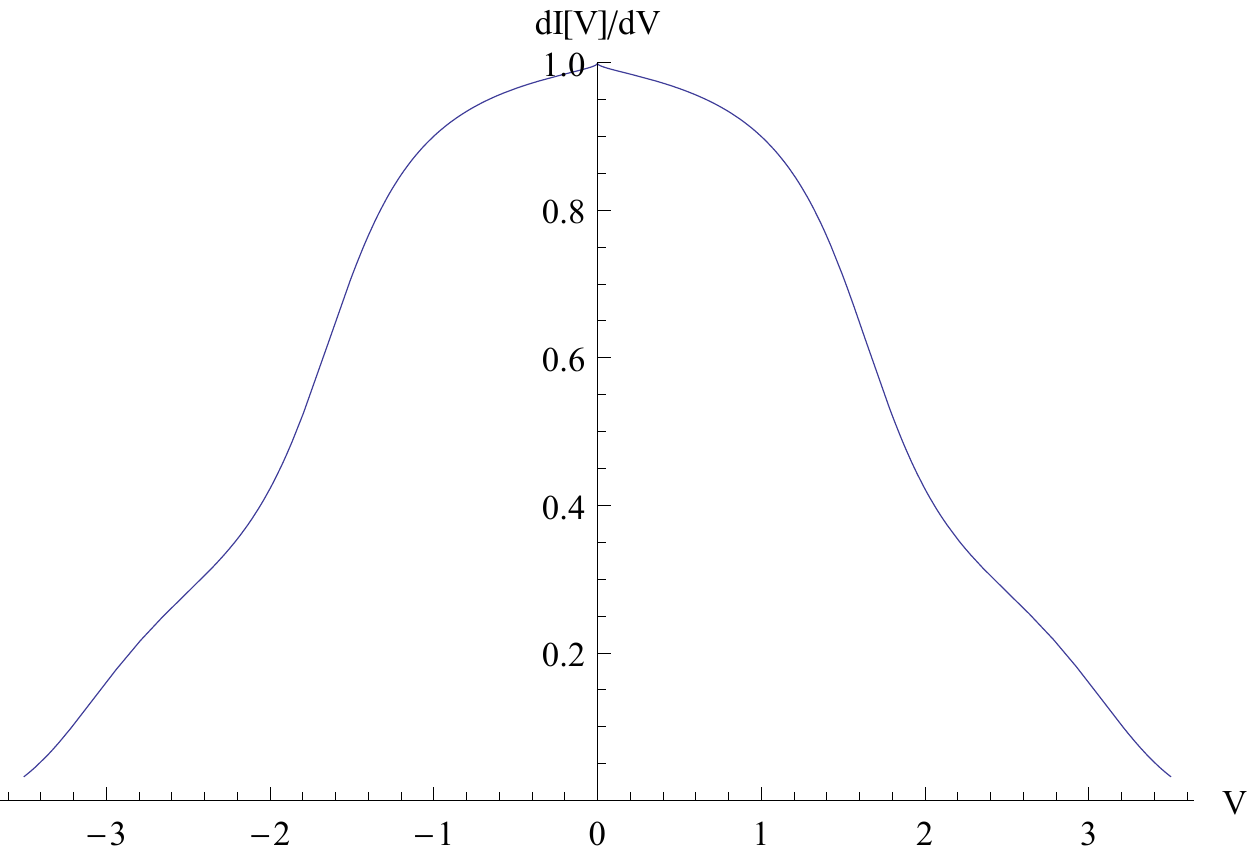} 
\end{center}
\caption{$\frac{dI}{dV} $  the differential conductivity for the Andreev crossed reflection}
\end{figure}

 \noindent
We find that  for a pair of vortices the Andreev crossed reflection obeys  $\frac{dI(V)}{dV}|_{V\rightarrow 0}  \longrightarrow \frac{e^2}{h}$  (see fig.$1$) and for a single vortex $\frac{dI(V)}{dV}  \rightarrow 0$.

\vspace{0.2 in}

\textbf{XII-Detection of the Majorana Fermions for two metallic rings  pierced by a magnetic flux which  by  $p-wave$}

\vspace{0.2 in}

We consider a situation where two metallic rings are attached to the   two ends of the  $ p-wave$  (which has  two zero modes at the of the wire, $\gamma_{1}$ and $\gamma_ {2}$) (see \cite{rings,davidMajorana}).

\noindent 
 We consider the special case where $ L=Nl$ ($L$ is the wire length and  $ l $   length of each  ring )
The flux in ring one is  $ f_{1}$ and in   ring  two is  $ f_{2}$. Using   periodic boundary  conditions  $ C_{1}(-l)=C_{1}(0)$   and  $ C_{2}(-l)=C_{2}(0)$  we perform a gauge transformation  $\hat{C}_{1}(x)=e^{i\frac{2\pi x}{l}f_{1}}C_{1}(x) $, $\hat{C}_{2}(x)=e^{i\frac{2\pi x}{l}f_{2}}C_{2}(x) $  which introduce   twisted  boundary conditions.  We find the spectrum  of the  uncoupled rings  as a function of the momentum $ \frac{2\pi}{l}n$, $ n=0,\pm1,\pm 2,...$.
\begin{equation}
H_{0}=\sum_{n}\frac{\hat{\gamma}}{2}\hat{C}_{1}^{\dagger}(n)(n+f_{1})^{2}\hat{C}_{1}(n) +\sum_{n}\frac{\hat{\gamma}}{2}\hat{C}_{2}^{\dagger}(n)(n+f_{2})^{2}\hat{C}_{2}(n)
\label{esp}
\end{equation}
$H^{Maorana}=\epsilon_{0}\zeta^{\dagger}\zeta $ is the Hamiltonian of the $p-wave$ wire restricted only to the Majorana modes. We express the Majorana fermions  in terms of the fermion fields $ \zeta$ and $\zeta^{\dagger}$, 
\begin{equation}
\gamma_{1}=\frac{1}{\sqrt{2}}\Big[\zeta^{\dagger}+\zeta\Big] ,\hspace{0.1 In} \gamma_{2}=\frac{1}{i \sqrt{2}}\Big[\zeta^{\dagger}-\zeta\Big]
\label{Maj}
\end{equation}
The coupling between the wire and the two  rings takes the form :
\begin{eqnarray}
&&H_{t}=\frac{g}{\sqrt{2}}\sum_{n}\Big[(\hat{C}_{1}^{\dagger}(n) (\zeta+\zeta^{\dagger})  +  (\zeta+\zeta^{\dagger}) \hat{C}_{1}(n))  \Big]+(-i)\frac{g}{\sqrt{2}}\sum_{n}\Big[(\hat{C}_{2}^{\dagger}(n) (\zeta-\zeta^{\dagger})  +  (\zeta-\zeta^{\dagger}) \hat{C}_{2}(n))  \Big]\nonumber\\&&
\end{eqnarray}
\noindent We  integrate the rings degree of freedom and obtain an effective Majorana  impurity Hamiltonian.
\begin{equation}
 H_{eff} =\hat{\zeta}^{*T}(\omega)  \Big[ M(\omega;f_{1},f_{2})\Big]\hat{\zeta}(\omega) ;  \hat{\zeta}^{*T}(\omega) =\Big[\zeta^{\dagger}(\omega),\zeta,(\omega),\zeta^{\dagger}(-\omega),\zeta(-\omega)
\Big]
\label{impurity}
\end{equation}
\noindent
$M(\omega)$ is a $ 4 \times 4$ matrix which depends on $\Delta$ and $\delta$ (the function obtained by integrating the ring degrees of freedom) .  $\Delta$ and $\delta$ are given by:

\noindent
$\Delta=\frac{g^2}{8}(\Delta^{(1)}+\Delta^{(2)})$, $\delta=\frac{g^2}{8}(\Delta^{(1)}+\Delta^{(2)})$,
$\Delta^{(1)}=\sum_{n}\Big[\frac{\omega-E_{1}(n) +ix}{(\omega-E_{1}(n))^2+x^2}\Big]$ , $\Delta^{(2)}=\sum_{n}\Big[\frac{\omega-E_{2}(n) +ix}{(\omega-E_{2}(n))^2+x^2}\Big]$ ,$x\rightarrow 0$.

\noindent
 The   matrix  $M(\omega)$ is given by:

\noindent 
$M_{1,1}(\omega)=  \omega-\epsilon_{0}-\Delta $, $M_{1,2}(\omega)=M_{1,3}(\omega)=0$,$ M_{1,4}(\omega)=-\delta$;
$M_{2,1}(\omega)=0 $, $M_{2,2}(\omega)=-(\omega+\epsilon_{0})+\Delta$, $M_{2,3}(\omega)=\delta$,  $M_{2,4}(\omega)=0$;
$M_{3,1}(\omega)=0$ ,$ M_{3,2}(\omega)=\delta$, $ M_{3,3}(\omega)=  -(\omega+\epsilon_{0})+\Delta$, $M_{3,4}(\omega)=0$;
$M_{4,1}(\omega)=\delta $,$M_{4,2}(\omega)=M_{4,3}(\omega)=0$, $M_{4,4}(\omega)= (\omega+\epsilon_{0})-\Delta$;
$E_{1}(n)=\frac{\kappa}{2}(n+f_{1})^{2}-\mu$;
$E_{2}(n)=\frac{\kappa}{2}(n+f_{2})^{2}-\mu$.

\noindent
We integrate the Majorana Fermion and obtain the exact partition function:
\begin{equation}
\mathbf{Z}(f_{1},f_{2})=Z(g=0;f_{1},f_{2})\cdot det[M(\omega;f_{1},f_{2})]
\label{uncoupled}
\end{equation}
\noindent $ Z(g=0)$ is the partition function for  two uncoupled rings. The effect of the coupling is controlled by the Majorana contribution $det[M(\omega;f_{1},f_{2})]$.
Therefore the current is given by $I_{1}(f_{1},f_{2})=\Big[ \frac{ d Log ( \mathbf{Z}(f_{1},f_{2}))}{df_{1}}\Big]$     ; $I_{2}(f_{1},f_{2})= \Big[ \frac{ d Log ( \mathbf{Z}(f_{1},f_{2}))}{df_{2}}\Big]$. 
Due to the multiplicative form of the partition function the current is a sum of two parts  $I_{i}(f_{i}; g=0)$ (i=1,2)  and a second   part   is determined  by  the matrix  $ M(\omega;f_{1},f_{2}) $   and  is given by    $\delta I_{i}(f_{1},f_{2})$ , $i=1,2$ .
The current in each ring  is given by, 
\begin{eqnarray}
&&I_{1} (f_{1},f_{2})=I_{1}(f_{1}; g=0)  + \delta I_{1}(f_{1},f_{2};M)\nonumber\\&& 
I_{2} (f_{1},f_{2})=I_{2}(f_{2}; g=0)  + \delta I_{2}(f_{1},f_{2};M)\nonumber\\&&
\end{eqnarray}

\noindent 
We investigate  the case of  equal fluxes, $ f_{1}=f_{2}=f$. Due to the fact that $ L=Nl$ the hoping matrix elements are real.  In particular the Majorana energy $\epsilon_{0}$ couple like a regular impurity to a set of states determined by the two rings . Effectively the integration of the electrons in the rings  renormalizes  the energy $\epsilon_{0}$ to   $\epsilon_{eff.}(\epsilon_{0},f)=\epsilon_{0}+\Sigma(f)$   where $\Sigma(f) $ is the shift in energy caused by the energy in the ring $E(n,f)$.


\begin{figure}
\begin{center}
\includegraphics[width=2.5 in ]{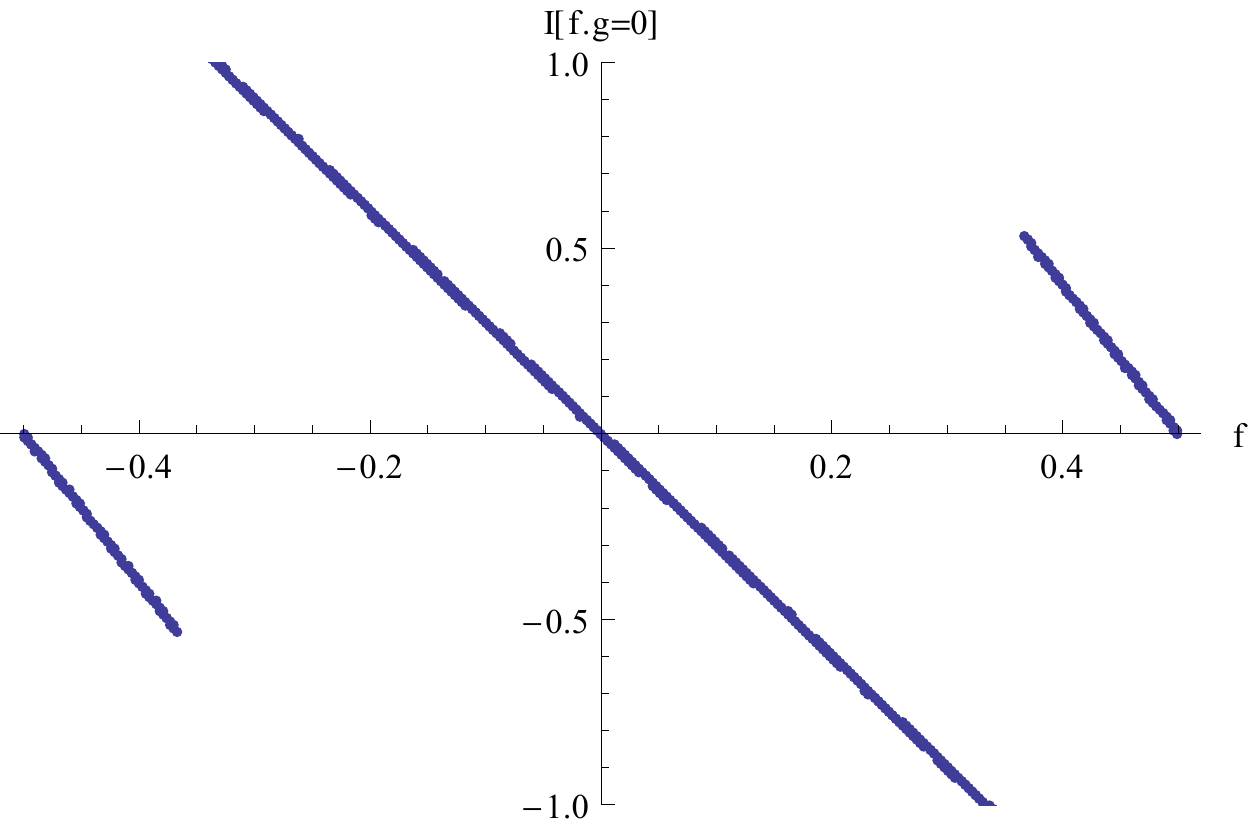}
\end{center}
\caption{The current for a single ring, the coupling constant to the wire is zero.  The current was computed for a fixed chemical potential,this explains the jump of the current  at $ f=\pm 0.4$.}
\end{figure}
\begin{figure}
\begin{center}
\includegraphics[width=2.5 in ]{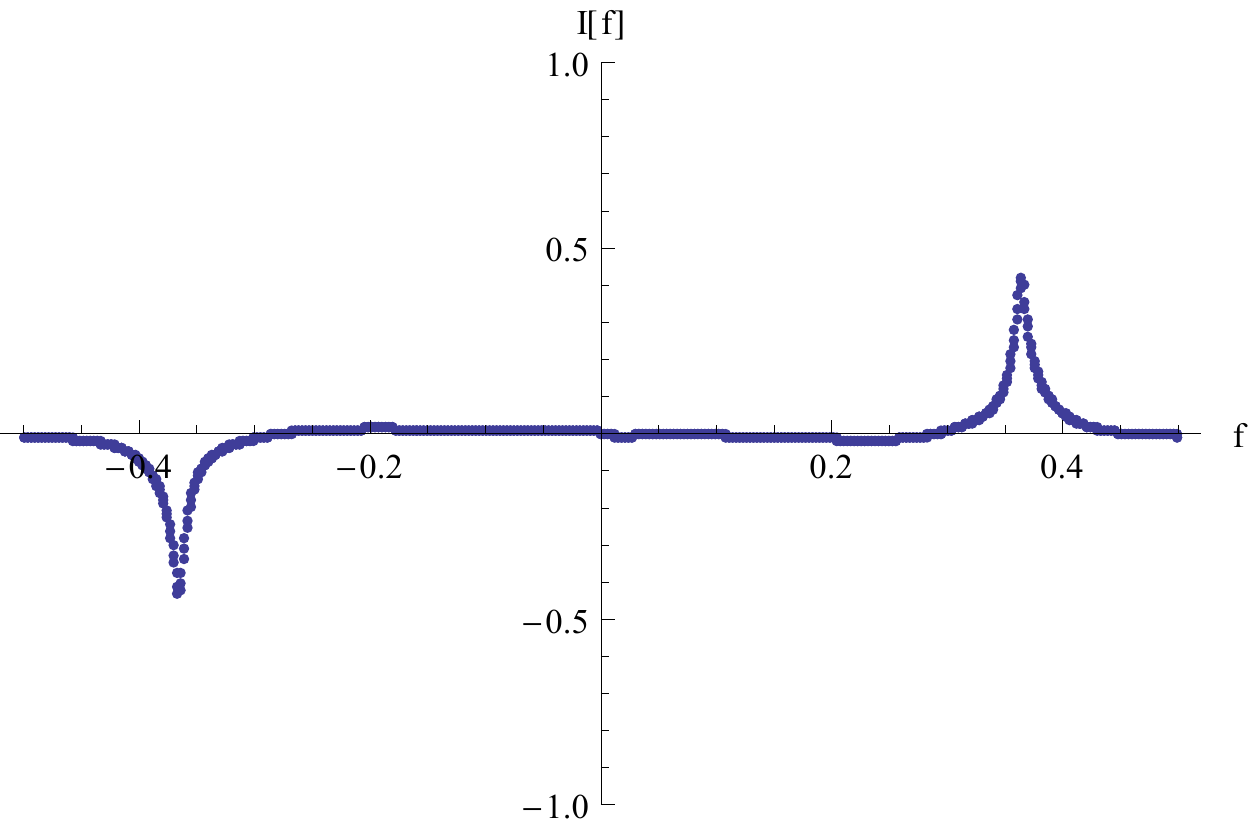}
\end{center}
\caption{The shift of the  persistent current  for the Majorana energy(two rings) $\epsilon_{0}=0.01$ .}
\end{figure}
\begin{figure}
\begin{center}
\includegraphics[width=2.5 in ]{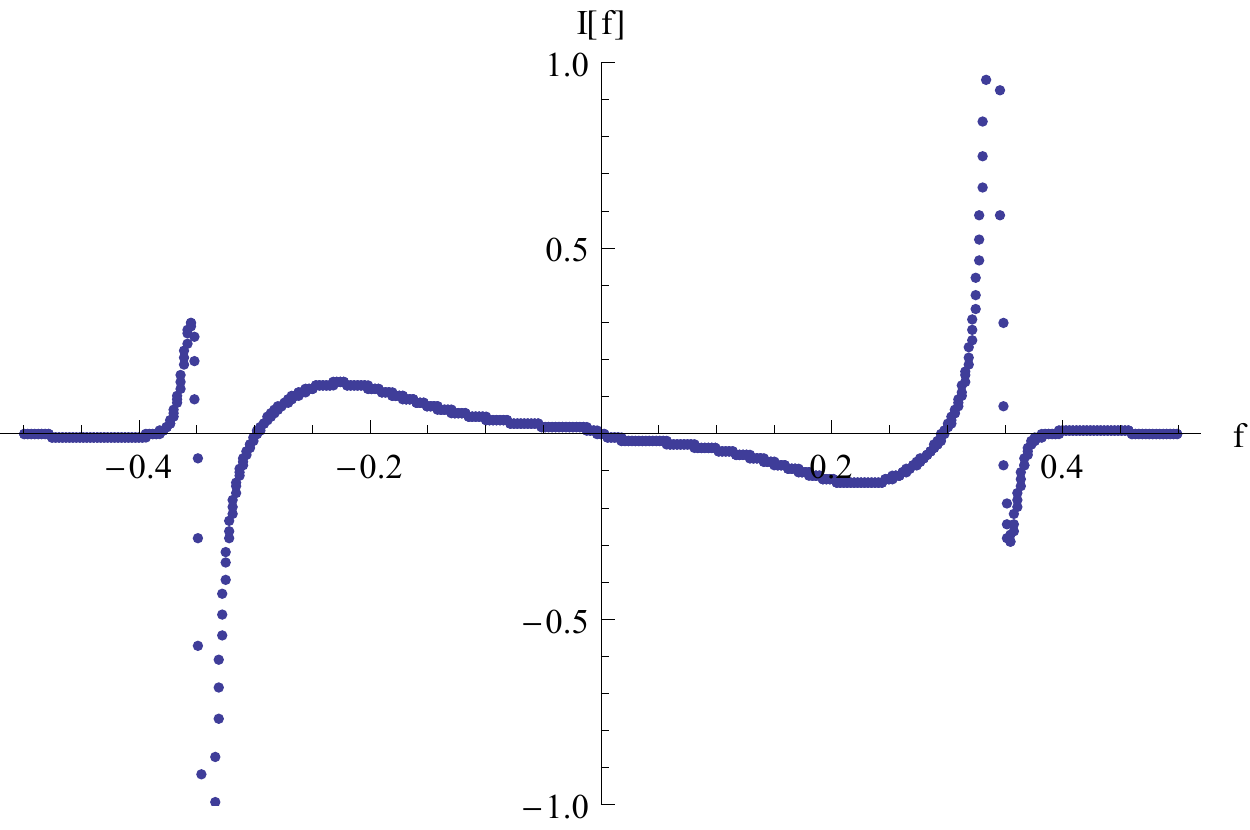}
\end{center}
\caption{The shift of the  persistent current  for the Majorana energy (two rings) $\epsilon_{0}=0.1$ .}
\end{figure}
 
This result show that when the Majorana energy goes to zero the only contribution to the persistent  current comes  from the perfect  wire in figure $2$ 
.
In the presence of a Majorana Fermion we have  in addition to  the persistent  current   from the perfect   wire  the  contributions   $\epsilon_{eff.}(\epsilon_{0},f)=\epsilon_{0}+\Sigma(f)$  determined by    the matrix   $M(\omega)$ given in fig.$3$ for the Majorana energy $\epsilon_{0}=0.01$ and in fig.$4$  for  the Majorana energy $\epsilon_{0}=0.1$.
\noindent
The persistent current is measured by scanning  with the \textbf{SQUID} \cite{Moler}  which measures the  change in the magnetization  (The magnetization is proportional  to the persistent current).  One subtract  from  the  value of the persistent  current, the current  from a single ring ,     and find  the dependence  of the persistent  current (the correlation part given  by   $\delta I_{i}(f_{1},f_{2})$ , $i=1,2$) on the Majorana energy. 
From the result obtained here it is suggested  that  for two \textbf{different fluxes},$f_{1}\neq f_{2}$ the effect of the fluxes will be local . The correlation between  the currents the two rings is negligible.

\noindent
 In the past only measurements of few rings was possible .
In recent years some of the experimental groups have claimed   to measure the persistent current  in a single ring.  Some of the complications  are  related to the fact that the measurements have been   done in diffusive limit (with 10-20 rings)  and not in the ballistic limit which is considered  in our calculation. 
The effect of the diffusive limit can be study by coupling the rings to a  noise bath.

\vspace{0.2 in}

 \textbf{XIII- Conclusions}

\vspace{0.2 in}

In conclusion the method of curved spaces introduced  in the context of the Spin Hall effect has been  applied to Topological Insulator and Weyl Semimetals. The  method consists on the fact that in momentum space the coordinates are given by a momentum derivative, due to the fact that the spinor changes in the B.Z. the coordinate becomes  covariant.  The appearance of the spin connections in the covariant derivative generates the curvature. The Topological invariants, the Chern number are obtained after we impose the  constraints 
 time reversal, or charge conjugation ( for Superconductors).
The Chern number are given by  the covariant coordinates commutators.
The formalism allows to obtain new Heisenberg equation of motion which depends on the momentum curvature and real space curvature (magnetic fields).
In particular the formalism is well suited to discuss $T.I.$  or  Weyl Semimetals on curved   surfaces. The Weyl Semimetals are characterized by monopoles and antimonopoles which can be investigated using a new set of complete eigenfunctions coined harmonic monopoles.

\noindent
For the  Topological Superconductors we considered  the effect of the Majorana fermions on coupled metallic  rings and computed the differential conductivity for the Andreev crossed reflection.

\vspace{0.2 in}

\textbf{Appendix A}

\vspace{0.2 in}

The physics   of  electrons in  a periodic crystal  is  determined   by the  eigenvectors (spinors)    $|U_{n}(\vec{k})\rangle$ ($n$ is the band-spin index) behavior   in  the  \textbf{ Brillouin Zone}  $\vec{k}\in T^{d}$ (torus in a $d$ dimensional  momentum space). This behavior  is similar   to  the parallel  transport  of a vector  around  a  curve.  We need to  find  the way the eigenvectors  change under transport in the Brillouin Zone  \cite{davidSpinorbit,Blount,Zak1,Zak2,Zak}.
The topological properties are encoded into  the   connection  $ \mathcal{A}_{i}$ (the vector potential in the momentum space) which measures  the changes of   $|U_{n}(\vec{k})\rangle$ when it  is transported  in the Brillouin Zone. The changes are   given by:

\noindent  $i d|U_{n}(\vec{k})\rangle-\Gamma^{m}_{i;n} dk^{i}| U_{m}(\vec{k})\rangle=0 $  (an index  which appears twice  implies  a summation ).  The matrix  $ \Gamma^{m}_{i;n}$ is given by
$ \Gamma^{m}_{i;n}\equiv
i\langle U_{n}(\vec{k})|\partial_{k^{i}}| U_{m}(\vec{k})\rangle=  \mathcal{A}^{(n,m)}_{i}(\vec{k})$  where   $\mathcal {A}$ is   the\textbf{ connection}.
Applying twice the (\textbf{exterior}) derivative  we define the \textbf{ curvature}   $ \mathcal{F}=d(d|U_{n}(\vec{k})\rangle $    ( see Eqs.$ 7.145  -  7.145$  , Nakahara (2008) page 285 \cite{Nakahara}) and find :

\noindent $\mathcal{F}=d(d|U_{n}(\vec{k})\rangle)=\Big[[\partial_{j}\Gamma^{l}_{i;n}(\vec{k})+\Gamma^{m}_{j;n}(\vec{k})\Gamma^{l}_{i;m}(\vec{k})]dk^{i}\wedge dk^{j}\Big]|U_{l}(\vec{k})\rangle=\frac{1}{2} [F_{i,j}]_{n,l}dk^{i}\wedge dk^{j}|U_{l}(\vec{k})\rangle$ ;  ( the symbol $\wedge$ represents the wedge product )

\noindent $ [F_{i,j}]_{n,m}=\partial_{i}\mathcal{A}^{(n,m)}_{j}(\vec{k})-\partial_{j}\mathcal{A}^{(n,m)}_{i}(\vec{k}) +i[\mathcal{A}^{(n,l)}_{i}(\vec{k}),\mathcal{A}^{(l,m)}_{j}(\vec{k})]$,

\noindent where $F_{i,j}$ is the matrix curvature with the matrix elements  $[F_{i,j}]_{n,m}$  given in terms of the commutator of the  \textbf{covariant derivative}  $\hat{R}_{i}=\hat{r}_{i}+\mathcal{A}^{(n,m)}_{i}$ , $\hat{r}_{i}=i\partial_{k^{i}}$ ; $ [\hat{R}_{i},\hat{R}_{j}]=F_{i,j}$.

\vspace{0.2 in}

\textbf{Appendix B}

\vspace{0.2 in}

\noindent  \textbf{ Topological invariant in  two space dimensions using the  topological  invariant in four space dimensions.}

\vspace{0.2 in}

The  topological response for time reversal invariant systems in one and two space dimensions is not entirely clear .  In three space dimensions we can use the   Chern-Simons form $ K_{3}(A,F)]$     to relate the the second Chern number $ C_{2}$ in  four space dimensions to three dimensions using the relation  $ch_{2}=d[K_{3}(A,F)]$.
In four dimensional  momentum space the second Chern number $ C_{2}$ is   given by an \textbf{ index  operator}.
In analogy with the index operator for the Dirac equation I introduce the index operator in the momentum space $Ind.\Big[i\mathbb{R}_{4}\Big]$:
\begin{equation}
 Ind.\Big[i\mathbb{R}_{4}\Big] =  Tr\Big[\gamma^{5 } e^{-\epsilon \int\,d^4k \bar{ C}(\vec{k})\Big( i\mathbb{R}_{4}(\vec{k})\Big)^2 C(\vec{k})}\Big] |_{\epsilon \rightarrow 0 }\longrightarrow  C_{2}
\label{Indexmiatrix}
\end{equation} 

\noindent The operator 
$i\mathbb{ R}$ is defined in terms of the non-Abelian  spin connection :
\begin{eqnarray}
 &&A^{\alpha,\beta}_{a}(\vec{k})\equiv  \langle U_{\alpha}(\vec{k})|i\partial_{a}|  U_{\beta}(\vec{k})\rangle\nonumber\\&&
i\mathbb{ R}_{4}(\vec{k})= i\sum_{a=1,2,3,4}(\gamma_{a}(x^{a}+A_{a}(\vec{k}))\equiv i\gamma_{a}X^{a}(\vec{k})\nonumber\\&&
 \gamma^{5 }=\left[\begin{array}{rrr}
                                       -1 & 0 \\
0 & 1 \\
\end{array}\right] \nonumber\\&&
\end{eqnarray}
$ \gamma^{5 }$ separates the conduction band from the valence band.

\noindent  In order to show that the index operator in four space dimensions  is related to the  index  operator in d=2 space  dimensions, we introduce the transformation :
\begin{equation} 
\mathcal{A}^{g(\theta)}(\vec{k})=g^{-1}(\theta,\vec{k})\Big(\mathcal{A}(\vec{k})+d\Big)g(\theta,\vec{k}) 
\label{conections}
\end{equation}
where $\theta$ parametrizes the $S^{1}$ circle (see  figure $13.4$  page $521$  \cite{Nakahara}). 
Next we construct a family of gauge fields: 
\begin{equation}
\mathcal{A}^{t,\theta}\equiv t \mathcal{A}^{g(\theta)};\hspace{0.1 in} 0\leq t \leq 1
\label{t}
\end{equation}

\noindent  The parameters $(t,\theta)$  form a disc $D^2$ with $\partial D^2=S^{1}$ .
We construct from$ D^2\times S^2$ a manifold  $S^2\times S^2$ . We will call the patch $(t,\theta) $ the northern hemisphere $U_{N}$ and  $(s,\theta)$  the south hemisphere $ U_{s}$ and the equator $S^{1}$ of $S^{2}$ corresponds to $t=s=1$ (see the  figure with the two half sphere,  figure $13.4$  page $521$  in \cite{Nakahara}).
The gauge potential in $2+2$ dimensions can be written as:
\begin{eqnarray}
&&\mathbb{A}_{N}(t,\theta,\vec{k})=\Big[0,0,\mathcal{A}^{t,\theta}(\vec{k})+g^{-1}(\theta,\vec{k})d_{\theta}g(\theta,\vec{k})\Big]\nonumber\\&&
\mathbb{A}_{N}(t,\theta,\vec{k})=\Big[0,0,\mathcal{A}(\vec{k})\Big]\nonumber\\&&
\end{eqnarray}
\noindent   $\mathcal{A}(\vec{k})$ is  the non-Abelian spin connection in two space dimensions. On the equator $t=s=1$ we have :
\begin{equation}
\mathbb{A}_{N}=g^{-1}\Big(\mathbb{A}_{s}+d+d_{\theta}+d_{t})g
\label{spin}
\end{equation}
 where $d$ is the \textbf{exterior derivative} \cite{Nakahara} in two dimensions, $d_{\theta}$ is the external derivative on $S^{1}$ and $ d_{t}g=0$. $\mathbb{A} =(\mathbb{A}_{N},\mathbb{A}_{S})$  is the spin connection in $2+2$ dimensions.
We use the relation,
\begin{equation}
 det\Big[i\hat{D}(\mathcal{A}^{g(\theta)})(\vec{k})\Big]\equiv\Big[\det(i\mathbb{R}(\mathcal{A}(\vec{k}))\Big]^{\frac{1}{2}}e^{iw(\mathcal{A},\theta)}
\label{det}
\end{equation}
 $|det[i\hat{D}]|$ is gauge invariant and only  $det[i\hat{D}]$ might have  anomalous behavior . On the boundary disc $\partial D^2=S^{1}$  the phase $e^{iw(\mathcal{A},\theta)}$ defines the maping  $\partial D^2\rightarrow S^{1}$
\begin{equation}
 i\hat{D}(A)= i\sum_{a=1,2,3,4}(\gamma_{a}(x^{a}+iA_{a}(\vec{k})P_{+}) ,\hspace{0.1 in} P_{+}=\frac{1}{2}(1+\gamma^{5 })
\label{anomalous}
\end{equation}
 $|det [i\hat{D}(A)]|$ is gauge invariant and only the phase  $ w(\mathcal{A},\theta)$ is anomalous and gives    the winding number $\nu_{1}$, on the disc $D^2$ there are points at which  $det [i\hat{D}(A)]$   vanishes.
\noindent  The  index  $Ind.\Big[i\mathbb{R}_{2+2}\Big]$ is given by,
\begin{equation}
Ind.\Big[i\mathbb{R}_{2+2}\Big]=\int_{S^2\times S^2} c_{2}(\mathbb{F})=\nu_{1}  
\label{nu}
\end{equation}
 where  the curvature $\mathbb{F}$ is given by,
\begin{equation}
 \mathbb{F}=(d+d_{\theta}+d_{t})\mathbb{A}+\mathbb{A}^2
\label{F}
\end{equation}
 $\nu_{1}$ is the winding number   which  is  is even  or odd and corresponds to the index $Z_{2}$ introduced earlier.

\noindent Following the procedure used before  which relates the Chern character  to the Chern-Simons form  $c_{2} = d[K_{3}(A,F)]$  we find:
\begin{equation}
\int_{S^2\times S^2} c_{2}(\mathbb{F})=\int_{D^2\times S^2}
 c_{2}(\mathbb{F}_{N})+\int_{D^2\times S^2}
 c_{2}(\mathbb{F}_{S})=\int_{S^{1}\times S^2}\Big[K_{2+1}(\mathbb{A}_{N},\mathbb{F}_{N})|_{t=1}-K_{2+1}(\mathbb{A}_{S},\mathbb{F}_{S})|_{s=1}\Big]
\label{procedure}
\end{equation}
Since $\int_{S^{1}\times S^2} K_{2+1}(\mathbb{A}_{S},\mathbb{F}_{S})=0$
we find:
\begin{equation}
Ind.\Big[i\mathbb{R}_{2+2}\Big]=\int_{S^{1}\times S^2}K_{2+1}(\mathbb{A}_{N},\mathbb{F}_{N})=\int_{S^{1}\times S^2}K_{2+1}(\mathcal{A}^{g(\theta)}+g^{-1}d_{\theta}g,\mathcal{F}^{g(\theta)})=-(\frac{1}{2\pi})^2 Tr[W d\mathcal{A}]
\label{result}
\end{equation}
\noindent where $ W\equiv g^{-1}(\theta)d_{\theta}g(\theta)$. For $\theta=0$  we find  $W=1$ and   $Ind\Big[i\mathbb{R}_{2+2}\Big]= \nu_{1}$   ($\nu_{1}$ is  the winding  number )  which is identical to the $Z_{2}$ index introduced by \cite{Kane}.

\noindent 
The procedure presented here  allows a direct construction of the $Z_{2}$ invariant in two dimensions as an emergent object from  four dimensions and therefore is related to the electromagnetic or sound wave response defined in four space dimensions. Contrary to early procedures which used dimensional reduction the  procedure proposed  is based on deforming  the spin connection $\mathcal{A}$ to a family of higher dimensions gauge potentials $\mathcal{A}^{g(\theta)}$.


\vspace{0.2 in}



\pagebreak


\end{document}